# Toward Software Measurement and Quality Analysis of MARF and GIPSY Case Studies – a Team 8 SOEN6611-S14 Project Report

Insight to the Quality Attributes achieved

Chanpreet Singh, Kanwaldeep Singh, Parth Manrao, Rashi Kapoor, Sagar Shukla, Shivam Patel, Simar Preet, Suman Alungh

Masters of Engineering in Software Engineering, Department of Computer Science and Software Engineering

Concordia University,

Montreal, Quebec, Canada

## I. ABSTRACT

Measurement is an important criterion to improve the performance of a product. This paper presents a comparative study involving measurements between two frameworks MARF and GIPSY. Initially it establishes a thorough understanding of these frameworks and their applications. MARF comprises of a number of algorithms for voice and speech processing etc. GIPSY on the contrary provides a multi lingual platform for developing compiler components. These frameworks are meant to provide an open source environment for the programmers or users and implement them in applications. Several metrics are used for object-oriented design quality assessment. We use these metrics to evaluate the code quality of both MARF and GIPSY. We describe how tools can be used to analyze these metric values and categorize the quality of the code as excellent or worse. Based on these values we interpret the results in terms of quality attributes achieved. Quantitative and qualitative analysis of metric values is made in this regard to elaborate the impact of design parameters on the quality of the code.

## II. INTRODUCTION

The intent of the paper is to describe and compare two frameworks MARF and GIPSY. It begins by firstly developing a deep understanding of the two frameworks. GIPSY (General Intentional Programming System) is a multi-intentional programming system which [15] serves as a platform for compiling and executing the programs written in Lucid Programming languages. MARF (Modular Audio Recognition Framework) is a Java based research platform which acts as a library in applications. Quality attributes of Object Oriented metrics such as MOOD, QMOOD, and Cohesion is analyzed and interpreted. In the latter section analyses is done using some well-known tools to cite differences in the quality of the code.

We uncover the internal and external quality characteristics of these frameworks. This paper also reviews code detectors for analyzing the code smells. In this work, we describe, analyze and evaluate the code quality to bring about relative comparisons between MARF and GIPSY.

## III. BACKGROUND

### A. OSS Case Studies

### A.A MARF

MARF is an open source, Java based research platform that is primarily used to measure and evaluate various pattern recognition algorithms. These natural language processing algorithms can be used for voice, sound, speech, text processing and even beyond. Technically, MARF is an assembly of pattern recognition algorithms, frameworks and API's that are put together to give a platform to researchers where they can measure and test new algorithms against each other. MARF follows pipeline architecture. The MARF framework is structured in such a way that it can be used for research and teaching in software engineering. The design patterns followed by the modules are Abstract Factory, Builder, Singleton, Visitor and Composite. The logical working of MARF is described as:

- Loading a sample (could be a WAV, SINE, TEXT, MP3 file).
- Preprocessing and normalizing the sample.
- Extracting its most distinguished features.
- Train the system or run classification.

The package structure of MARF is defined as:

- **MARF root package** – This package consists of Configuration and Version class.
- **Preprocessing** – It is the second stage in the pipeline and contains all preprocessing algorithms like n-pass FFT or CFE filters, normalization.
- **Feature Extraction** - Third stage of the pipeline implementations include FFT and LPC algorithms.
- **Classification** - Final stage in the pipeline. It includes distance classifiers and artificial neural network algorithms.
- **Statistics and Mathematics** - This can be used at any stage for statistical gathering and plain mathematics.
- **NLP** - It involves parsing of the natural language, grammar compilation, stemming etc.
- **Net** - For distributed MARF components.
- **Storage** - For storage management in MARF
- **Util** - General purpose utility modules
- **GUI** - General purpose storage modules [13][4]

Firstly, we discuss different aspects involved in the package structure of MARF.

To proceed with we start with algorithms. It covers some major algorithms like Fast Fourier Transform (FFT), Linear Predictive Coding (LPC), Artificial Neural Network, Cosine Similarity Measure, Pipeline's algorithm, Continuous Fraction Expansions (CFE) etc. The most accurate configurations were found to be for cosine similarity. [2] The NLP package includes algorithm for parsing, stemming and n-gram models. Along with these algorithms, MARF comes with some inbuilt utility modules that have nothing to do with NLP or pattern recognition as such. These utility modules provide thread management, debugging and logging, option processing tools, comparators common storage data structure for sorting. The pipeline has also been implemented in an extended way to run on different computers and nodes over SNMP to test its properties on distributed systems. [13]

Classical NLP components enlarge the pipeline architecture. The NLP statistical estimator's pipeline is used for smoothing estimators. These estimators act as classifiers. In this case omitted slow algorithm is not tested and debugged because this algorithm is very slow. There is one classifier known as zipfs law which exists in NLP pipeline and classical pipeline. This classifier is very slow in nature and is used to collect the most discriminated terms in the NLP pipeline.[10]

MARF also implements some interfaces that are made general because the pipeline does not know which modules are being called. The major interfaces used in MARF are IStorageManager, IDatabase, IPreprocessing, IFilter, IFeatureExtraction, IClassification. The algorithms used in MARF are not just limited to audio processing; they extend to all pattern recognition applications. This is an added feature of MARF. Certain quality attributes are inherent to architecture of MARF i.e. extensibility, modifiability, testability, maintainability, adaptability, compatibility, configurability and efficiency.[13][4]

To differentiate between the best algorithm mean and median cluster approach is used for speech processing tasks. The statistics is used to analyze the performance of the task of text independent speaker identification, their gender and their accent in order to select the best combinations of the preprocessing, feature extraction, and classification algorithms in the MARF's application, SpeakerIdentApp. One of the main goals is to provide a tool for comparison of algorithm implementations in the homogenous environment and to allow module selection based on the configurations options supplied by applications. SpeakerIdentApp is the application of MARF which is used as an open source and has number of command line options for scripting purposes to allow automated testing. It requires the presence of disjoints, sets of training, testing sample files and their IDs for validation and verification purposes during experiments. The previous studies conducted using mean clusters of features vectors were incomplete as those results were not the best combinations. This leads to proposed solution of selecting the best combination using median cluster vectors. [2]

MARF is used as a tool to recognize patterns to provide comparative environment for numerous algorithm. In order to achieve the quality attributes like extendibility, interfaces and data structures need to be defined to encapsulate and store the data for comparing. Set of methods and tools were employed in the MARF approach which resulted in different results. Some techniques were based on text independent speaker identification using mean and median cluster, gender identification etc. while the others were based on writer identification. All these experiments resulted in top, intermediate and worst configurations. For instance, median clusters can yield a top configuration while the configuration that was top for the mean was no longer accurate.

In a nutshell, MARF provides a pattern detecting capability for various learning and recognizing tasks and the same was illustrated by means of the experiment. It provides uniformity, consistency, precision in results and extensibility using plug ins support but at the same time it becomes tedious when the number of algorithms are large.[16]

Next, we proceed with analyses of speech reorganization task susing MARF. The core part of study is to identify the speakers their accent and their gender via machine learning. This study is advancement of text independent speaker identification. The statistic of the reorganization accuracy is used to choose effective combination of preprocessing, feature extraction, categorizing algorithm in the MARF and speaker identification application. The amalgamation of algorithm in the pattern reorganization for the speaker identification is not effective all the time when it considers attributes like accent and gender. The procedure for pattern recognition is done by loading the sample, extracting the noise, identifies the most notable characteristic, and then instructing the system to recognize the speakers or subject. Output of this is 32-bit distinct integer number, which points out who are the speakers. Default parameter is used for each concrete module. As a result and analysis, speaker identification application gathers the statistics for effective and ineffective solutions. Algorithm is chosen based on three characteristic like gender identification, speaker identification and accent identification. Each option has some parameters for the application, which choose the accurate algorithm at a run time. It helps to identify attributes of various speakers, which is useful for safety and security. Applications and frameworks are effective and efficient way to examine the algorithm for different attributes. [7]

The classical audio and text signal processing techniques were used to achieve a higher efficiency in writer identification tasks. Many pattern recognition algorithms were built around MARF framework to do a comparative study. Current writer identification techniques rely on classical methods like skeletonizing; contouring and then the results are compared with trained data set. These techniques were highly accurate but are also time consuming during bulk processing. The proposed solution advised to treat sample pages as 1D or 2D arrays of data and then loading and filtering the same and treating each hand-written image sample as a wave form then we continue the classical feature extraction, training and classification tasks using a comprehensive algorithm set within MARF's implementation. ID array were used as storage because it is the baseline storage mechanism of MARF and it is less storage consuming and achieve high accuracy in writer identification task. MARF was designed to act as a platform to test, try and compare various algorithms found in industry for sample loading, preprocessing, feature extraction, training and classification tasks. MARF is also designed to be very configurable .To enable the experiments in this work and their results we required to do the alteration of the MARF's pipeline through its plug-in architecture which shows MARF's extensible architecture. Initial loading plays a vital role in an effective outcome. The experiment was made to run on two different hardware setups which showed that performance was improved if hardware used was stronger and powerful. The results showed an increase in efficiency which comes at the cost of dip in accuracy (approximately 20% accuracy). Certain quality attributes that are inherent to MARF's design is that it is extensible, modifiable, testable, maintainable, adaptable, compatible, configurable and efficient. [21]

Due to security reasons measures were developed which are intended to address the security concerns which arises when the system communicates in or with a system in an untrusted local network. Though the overhead of any security measure is not mandatory unless the connection is from unsecured or untrusted network, thus we need an optional security measure which can be called in when required. The GICF framework was introduced to ensure more operability between the intentional and imperative programming language, but this leads to security violations when embedding a potential vulnerable unsigned code from a untrusted location. The DMF (Demand Migration Framework) architecture was introduced to establish communication after ensuring the demand source. The SNMP was introduced in the DMARF for further managing its nodes from the security perspective. Though both systems GIPSY and MARF have different execution models demand driven and pipelined for the later. The risk of injecting malicious errors is there in both the systems at different stages of execution of data. The JDSF was introduced to lessen the security threats. The main aim was to achieve confidentiality, Integrity, Authentication and Availability of the data. The JDSF was used as an upper layer on the two systems, GIPSY and DMARF, to ensure system security is maintained. However, the security issues such as confidentiality, Integrity, and Authentication issues were well addressed by the JDSF, It lacks in ensuring the availability issue as availability can be due to malicious code having infinite loops that have to be either aborted or disallowed as far as static code analysis allows. [18]

The above applications of MARF provide sufficient basis to achieve quality in terms of non-functional requirements termed as Quality attributes.

The main aim of introducing JDSF was to achieve confidentiality, Integrity, Authentication and Availability of the data. Integrity and confidentiality aims at building a secure environment against the unwanted errors introduced while its execution.

Apart from the security, the applications also provide high level of reusability as the same source code can be used with different plug in support and add ons to generate new features. It thereby, helps in reducing the implementation time as the same system can be expanded to get new features.

The application for combining and comparing multiple algorithms provides the ease of comfort in terms of its usage.

These measures when adhered resulted in the overall system's productivity and efficiency at each development phase.

Although MARF offers features but it also demands changes. The conventional pipeline architecture of MARF offers no concurrency which resulted in the requirement of distributed MARF known as DMARF.

*A.B   GIPSY*

General Intentional Programming System (GIPSY) is a multi-language intentional programming system that investigates intentional programming. Intentional programming refers to the concepts that enable a source code to reflect the intention of the programmers while they conceive their work. It is an ongoing effort for development of efficient multi lingual programming language development framework. GIPSY framework is used for developing compiler components for other languages of intentional nature, and to execute them on a language-independent run-time system. [6], [14]

Lucid is an example of multidimensional intentional programming language, which is successfully applied to resolve the problem and understand the nature of the problem. In Lucid, programs can be evaluated in parallel. However in order to profitably exploit parallelism, GLU was introduced (Granular Lucid). But, as the syntax was modified in the following years of Lucid language, GLU complier was not able to adapt the new changes in the language. With the idea of GLU in mind, a new system was made with similar capabilities but is more flexible and is known as GIPSY. Three solutions which intend to showcase the usability of the proposed GIPSY model. Firstly, GIPC (General intentional programming language compiler); Secondly, GEE (General education engine); and lastly, RIPE (Intentional Run time programming environment).

The GIPSY model was designed to meet the following qualities:

- Language independence: execution of program written in any language of Lucid family.
- Scalability: to make efficient use of distributed computing.
- Flexibility of execution architecture: to make execution architecture configuration changes during execution.
- Opacity of run time consideration: executions considerations expressed into the source programs.
- Observable execution structure: provides infrastructure to observe all execution components. 0, [6], [11]

This new system is able to cope up with the diversity of Lucid family of programming languages. Programs written in many flavors of Lucid can be compiled and executed. However, GIPSY is a complex system, whose architecture permits high scalability; it still does not have the self-management capacities to achieve it. Previous GIPSY network was managed by command line. To subdue this problem of complexity, GIPSY has to be converted into a self-adaptive, autonomic computing system. For this the proposed solution is to implement the graph base GUI to manage the runtime system component of GIPSY effectively and efficiently. GIPSY network is represented by the node and connection between them. The preconfigured files are used to initialize the GIPSY network, which can be loaded at runtime. The architectural design of autonomic version of GIPSY (AGIPSY) is aimed at presenting GIPSY as an autonomous system that works in a much simpler and smarter way.

The proposed architecture is intended to automate the existing GIPSY. It is based on multiple interacting GIPSY nodes that are designed as autonomic elements by applying the AE architecture for ASSL

AGIPSY implements the self- monitoring, a core feature of AC. It enables automated monitoring and management of complex multi-tiered, distributed workloads across GIPSY infrastructure to better achieve goals for end user services.

The GIPSY architecture is multi-tiered with the programs divided into four tasks assigned to separate tiers. The four tiers are namely:

- **GIPSY Manager Tier (GMT) -** It enable the registration of GIPSY nodes (GNs) to a GIPSY instance (a set of interconnected GIPSY tiers), and the allocation of various GIPSY tiers to these nodes
- **Demand Worker Tier (DWT) -** It exposes a set of demand workers (DWs) that can process procedural demands from the intentional program.
- **Demand Generator Tier (DGT) -** It encapsulate a demand generator (DG), which generates demands using an Intentional Demand Processor that implements the eductive model of computation
- **Demand Store Tier (DST) -** It is a communication system that connects the other GIPSY tiers by migrating demands among them, by connecting them to the Demand Migration System (DMS).

The user has the following advantages:

- User has the functionality to manage the GIPSY network manually.
- It allows user to depict, envisage and control the whole GIPSY network and interconnection between nodes at runtime using graph.
- It also allows the user to operate the whole GIPSY network by converting simple graphical user interface into complex system, which performs all operation without minimal intervention of user.

Challenges Faced:

- Introduction of a simulator into the NM architecture.
- Recovery strategy at both AGIPSY and GN levels and the GN healthy status expression.
0, [8], [14], [22]

With an aim to have higher reliability and throughput while executing a request, two systems were proposed: JINI and JMS DMS. Though both the methodologies have common communication medium, the Javaspace, the architecture and operation, however, is different for both:

- JINI has higher reliability but due to its memory restrictions, it is not reliable and has lesser availability. It often crashes when high memory usage is there whereas the JMS DMS has a better reliability but the throughput is less in comparison to the JINI.

- The JINI has the ability to work with pending data packets which is not there in JMS and the JMS has a unique producer/consumer models which defines the sender and recipient of the message. [19]

GIPSY provides a platform to investigate into intentional and hybrid intentional-imperative programming. Lucid (a family of intentional programming language) is translated to GIPL (Generic Intentional Programming Language) on which GIPC is based. Generic language also solved the problem of language dependence of run time system by allowing common representation of all compiled programs. Advancements were proposed in the design and implementation of multi-tier run time system of GIPSY:

A generic distributed run time system has been proposed. Certain work has been done in this direction using Java RMI and some implementations were based on Jini and JMS, which are lacking clarity and are complicated. A solution proposed is to extend the work on GLU's generator –worker architecture to be multi-tier as it was not much scalable and flexible. In the proposed solution all four implemented distributed computation prototypes i.e. multi-threaded, RMI, Jini and JMS will be integrated by applying abstract factory methods and strategy design patterns to make it more extensible and maintainable. To overcome GLU's inflexibility GIPSY was designed to include a generic and technology independent collection of Demand Migration Systems which implement the Demand Migration Framework (DMF) for a particular technology or technologies to communicate and store Information. [22]

A general type system was developed for static and dynamic type checking to be used by GIPSY at compile and run time respectively. Since, the type system was stringent before, therefore, by deploying the type system in GIPSY any Lucid dialect became capable of calling functions or methods written in imperative language and perform the required type conversion or semantic analysis if needed. It enables programmer to declare type of variables and return values for both intentional and imperative functions as a binding contract between inter language invocations despite the fact Lucid is not explicitly typed.

The new type system on implementation offers flexibility, adaptability, and extendibility in multiple programming language frameworks and above all it provides ease of use. Lucid method becomes capable of calling Java methods. For each return type in Java and expression type in Lucid, a corresponding GIPSY type is used. This flexibility is attained in GIPSY using STT (simple theory of types).

This is a crucial quality attribute, which was not there in the previous system. Flexibility provides adaptability of the system with multiple programming languages. [15]

*A.C Summary*

In this section we compute the following values from the application source code of MARF and GIPSY:

- Number of Files.
- Number of Classes.
- Lines of Text.
- Programming languages.

We have used Code Pro Analytix, Logiscope and UNIX shell scripting and commands in this regard. In the following section we discuss the methodology used to obtain these results.

Code Pro Analytix is a Java software measurement and testing tool for Eclipse developers inclined towards improving software quality and reducing development costs and schedules. On selecting Compute Measures from the menu item, the default measures set will be executed which produces the metric result set which can viewed using measures view. The result is in a tabular form described below with its first column providing a list of measures and the second column representing the value obtained.

The below results are extracted from the MARF and GIPSY source code in a tabular format.

Table 1: MARF Analysis using Analytix

| Name | Value |
|---|---|
| interface | 16 |
|    public | 16 |
|    protected | 0 |
|    package | 0 |
|    private | 0 |
| class | 200 |
|    public | 197 |
|    protected | 2 |
|    package | 0 |
|    private | 1 |

Table 1 gives information about the number of classes. MARF consists of a total of 200 classes and the public, protected and private bifurcation of classes is illustrated in table above.

Table 2: GIPSY Analysis using Analytix

| Name | Value |
|---|---|
| interface | 75 |
|    public | 72 |
|    protected | 0 |
|    package | 3 |
|    private | 0 |
| class | 626 |
|    public | 540 |
|    protected | 0 |
|    package | 65 |
|    private | 21 |

The above table gives information about the number of classes for GIPSY. From the above output it is evident that GIPSY has in total 626 files, out of which public and private are 540 and 21 respectively.

Thus we conclude that GIPSY is more complex as compared to MARF because higher the number of methods and classes higher is the complexity.

Another tool used for the computation of these values is Logiscope. It is software measurement tool which is capable of detecting coding defects at an early phase of the development life cycle thus reducing correction costs. We have used this tool to measure quality of Modular Audio Recognition Framework (MARF) and General Intensional Programming System (GIPSY). Based on the output of this tool we found below results:

Table 3: MARF Analysis using Logiscope

| Metric Name | Min | Max | Value |
|---|---|---|---|
| Application Comment Rate | 0.20 | +∞ | 0.41 |
| Application effective lines of code | 0 | 200000 | 16029 |
| Number of application functions | 0 | 3000 | 2419 |
| Depth of inheritance tree | 1 | 5 | 8 |
| Number of Lines of comments | -∞ | +∞ | 21436 |
| Number of lines | 0 | 300000 | 52633 |
| Number of lines of code | -∞ | +∞ | 24479 |
| Number of lines with lone braces | -∞ | +∞ | 8450 |
| Number of statements | 0 | 100000 | 9845 |
| Sum of cyclomatic numbers of application functions | 0 | 6000 | 3762 |
| Average complexity of functions | 1.00 | 3.00 | 1.56 |

Table 4: GIPSYAnalysis using Logiscope

| Metric Name | Min | Max | Value |
|---|---|---|---|
| Application Comment Rate | 0.20 | +∞ | 0.16 |
| Application effective lines of code | 0 | 200000 | 79836 |
| Number of application functions | 0 | 3000 | 6447 |
| Depth of inheritance tree | 1 | 5 | 7 |
| Number of Lines of comments | -∞ | +∞ | 22262 |
| Number of lines | 0 | 300000 | 139680 |
| Number of lines of code | -∞ | +∞ | 104083 |
| Number of lines with lone braces | -∞ | +∞ | 24247 |
| Number of statements | 0 | 100000 | 59862 |
| Sum of cyclomatic numbers of application functions | 0 | 6000 | 22704 |
| Average complexity of functions | 1.00 | 3.00 | 3.52 |

Lastly, we created a script in Linux OS hosted locally on Macbook 13.1.0 version xnu-2422.90.20~2 / RELEASE X86_64 to traverse the directory of the source code to obtain the values of the above fields. The script uses find command to search all files by specifying the type as f for files in the search criteria and then taking its count using wc.
For finding the number of classes all files with .class extensions is counted after compiling the code using Javac utility. For finding the lines of text, all the file names were redirected in a single file using awk command and were executed in a loop to count the number of lines. The below screenshots gives the output that script generates on execution:

The below output was generated for MARF only using script.

```
******************* This script calculates the following values for Java Code *******************
Number of files in the source code are :     546
Number of class files in the source code are :     186
Number of lines of text are :    57821
```

Figure 1: MARF Analysis using Linux Shell Scripting

Please note that the value 186 here only refers to the class files not the total number of files. Finally in order to determine the programming languages we installed Cloc package i.e. cloc-1.60.tar.gz on the Linux machine. After, successful installation of the package, Cloc utility was used to determine the values.

```
Chanpreets-Macbook:cloc-1.60 root# ./cloc /Users/apple/Desktop/marf
    608 text files.
    363 unique files.
    300 files ignored.

http://cloc.sourceforge.net v 1.60  T=3.91 s (65.9 files/s, 15324.9 lines/s)
-------------------------------------------------------------------------------
Language             files          blank        comment           code
-------------------------------------------------------------------------------
Java                   201           6762          21421          24450
Perl                     3            499            830           2534
make                    43            538            345           1509
XML                      3              3            105            392
Bourne Shell             5             50             64            168
CSS                      1             25             11             94
Bourne Again Shell       1             16             36             36
Ant                      1             14             51             17
-------------------------------------------------------------------------------
SUM:                   258           7907          22863          29200
```

Figure 2: Cloc for MARF

```
Chanpreets-Macbook:cloc-1.60 root# ./cloc /Users/apple/Desktop/proj/gypsy
    1993 text files.
    1293 unique files.
    1352 files ignored.

http://cloc.sourceforge.net v 1.60  T=11.86 s (76.8 files/s, 13428.7 lines/s)
-------------------------------------------------------------------------------
Language             files          blank        comment           code
-------------------------------------------------------------------------------
Java                   598          14460          21177         104024
ASP.Net                 79           1055              6           3287
make                   100           1280            617           2840
HTML                    40            717           1102           1650
C/C++ Header            19            305           1707            826
C                        5            208            324            797
Bourne Shell            16            127            210            439
DOS Batch               33            205             32            361
XML                      7            130            318            271
Visual Basic             1             62             94            131
Ant                      2             18              7             98
Bourne Again Shell       8             46             34             86
CSS                      2             17             44             74
Haskell                  1             10              1             44
-------------------------------------------------------------------------------
SUM:                   911          18640          25673         114928
```

Figure 3: Cloc for GIPSY

Finally, to summarize all the findings cumulatively:

**MARF Consolidated Data:**

| Metric | Value |
|---|---|
| Lines of Comment | 22863 |
| Number of Lines | 139680 |
| Number of Lines of Code | 29200 |
| Number of Classes | 200 |
| Public Classes | 197 |
| Protected Classes | 2 |
| Private Classes | 1 |
| Languages | 8 |
| Total Number of Files | 258 |

**Language Percentage Level: MARF**

| Language Components | File Utilization % |
|---|---|
| Java | 77.90 |
| Perl | 1.6 |
| Make | 16.66 |
| XML | 1.6 |
| Bourne Shell | 1.9 |
| CSS | 0.38 |
| Bourne Again Shell | 0.38 |
| Ant | 0.38 |

**GIPSY Consolidated Data:**

| Metric | Value |
|---|---|
| Lines of Comment | 25673 |
| Number of Lines | 52633 |
| Number of Lines of Code | 114928 |
| Number of Classes | 626 |
| Public Classes | 540 |
| Protected Classes | 0 |
| Private Classes | 21 |
| Languages | 14 |
| Total Number of Files | 911 |

**Language Percentage Level: GIPSY**

| Language Components | File Utilization % |
|---|---|
| Java | 65.64 |
| ASP.net | 8.67 |
| Make | 10.9 |
| HTML | 4.39 |
| C/C++ Header | 2.08 |
| C | 0.54 |
| Bourne Shell | 1.75 |
| DOS Batch | 3.62 |
| XML | 0.76 |
| Visual Basic | 0.21 |
| Ant | 0.10 |

| Bourne Again Shell | 0.87 |
| CSS | 0.21 |
| Haskell | 0.10 |

By considering above results it is evident that GIPSY is much complex than MARF if we consider LOC as a measure.

*B. Metrics*

*B.A  A hierarchical model for object-oriented design quality assessment.*

There is a continuous increasing dependence of our society on software and we demand for a consistent and error free operation of the software system which intensifies the demand for quality software. There is a need of metrics and models which can be applied in early stages of development and help improving quality of software and reduce rework later on. The new model presented has low level designed metrics and quality is assessed as an aggregation of the model's individual high-level quality attributes.

The QMOOD development extends from Dromey's Quality framework which is based on 3 principles: product properties that influence quality, a set of high level quality attributes and a means of linking them. The figure below shows the four levels and three mappings used in QMOOD.

First Level    Second Level    Third Level    Fourth Level

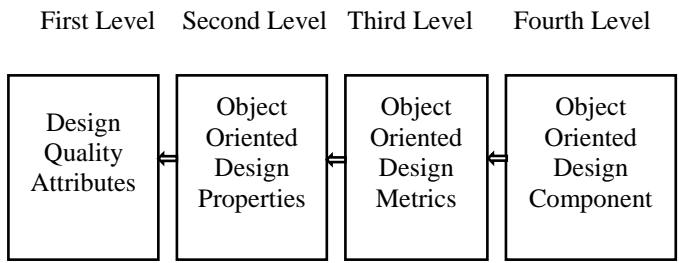

**Figure 4: Level Mapping in QMOOD [23]**

Design properties can be assessed by examining the internal and external structure, relationship, and functionality of the design components, attributes, methods and classes. There are several design metric available for design properties like inheritance, abstraction. But for some like encapsulation, composition etc. no design metric exists. Metrics to calculate coupling, cohesion and complexity exists but require implementation of classes before they are calculated. Hence they are of no use in QMOOD. Hence five new metrics were defined i.e. DAM, DCC, CAM, FMOA and MFA. Object Oriented Design Components is a set of components which can help analyze, represent and implement an object oriented design which should include attributes, methods, objects, relationships and class hierarchies. An automated tool QMOOD++ collects metric data from the components of a design. Quality carrying properties identified for attributes, methods and class can be grouped into set of design properties as shown in table below.

**Table 5: Definitions of Quality Attributes [23]**

| Quality Attribute | Definition |
|---|---|
| Reusability | Characteristics that allow to reuse a design to a new problem without much effort |
| Flexibility | Characteristic that allow to adapt changes in a design |
| Understandability | Characteristic of a design that enable it to be easily understand and comprehend |
| Functionality | Responsibilities assigned to a class are made available by classes through public interface |
| Extendibility | Characteristic of a design to allow to incorporate new requirements in the design |
| Effectiveness | Design's ability to achieve the desired functionality |

Table below shows the Index Computation Equation for the attributes:

**Table 6: Index Computation Equations [23]**

| Quality Attribute | Index Computation Equation |
|---|---|
| Reusability | -0.25*Coupling+0.25*Cohesion+0.5*Messaging+0.5*Design Size |
| Flexibility | 0.25*Encapsulation-0.25*Coupling+0.5*Composition+0.5*Polymorphism |
| Understandability | 0.33*Abstraction+0.33*Encapsulation-0.33*Coupling+0.33*Cohesion-0.33*Polymorphism-0.33*Complexity-0.33*Design Size |
| Functionality | 0.12*Cohesion+0.22*Polymorphism+0.22*Messaging+0.22*Design Size+0.22*Hierarchies |
| Extendibility | 0.5*Abstraction-0.5*Coupling+0.5*Inheritance+0.5*Polymorphism |
| Effectiveness | 0.2*Abstarction+0.2*Encapsulation+0.2*Composition+0.2*Inheritance+0.2*Polymorphism |

To validate QMOOD quality model two popular window frameworks were chosen i.e. MFC and OWL so that comparison is done between designs that were developed for similar requirements and objectives. Various versions of both frameworks were analyzed by QMOOD++ and through human evaluator. For both cases it showed that the value of each quality attribute improves for new system release for both frameworks. Total Quality Index (TQI) is obtained by summing the values obtained for six quality attributes which was used to rank project designs. Spearman's correlation coefficient, Rs, was used to test the correlation between QMOOD's and human evaluator assessment.

Rs is calculated using below formula:

$$rs = 1 - \frac{6\sum d^2}{n(n^2-1)} \qquad 1$$

where, $-1.00 <= rs <= +1.00$

There exists a null hypothesis and an alternative hypothesis which says zero correlation and significant positive correlation between relative rankings of design's quality indicator evaluated using QMOOD and human evaluator. This technique indicates that correlation can be effectively used in monitoring the quality of software product. [23]

*B.B Validation of Object-Oriented Design Metrics as Quality Indicators*

The case study presents the results of the empirical investigations conducted on the object oriented design metrics. The study is based on finding quality indicators mainly fault proneness of object oriented classes. Based on the findings, we can measure the fault proneness of classes on internal factors (e.g. size and cohesion) and external factors (e.g. coupling). The study is conducted on eight independent medium sized management systems.

Recently, there has been a significant increase in the implementation of object oriented paradigm in the software development field. Thus it is important to validate the metrics that will be used to test the quality of the object oriented paradigm. The probability of fault detection is the most straight forward and practical measure of fault proneness. Six Chidamber and Kemerer's metrics have been used to measure the fault proneness of OO classes. These are: Weighted methods per class (WMC), Depth of Inheritance Tree of a class (DIT), Number of children of a class (NOC), Coupling between objects (CBO), Lack of Cohesion on Methods (LCOM) and Response for a Class (RFC).

Some of the hypothesis for testing each metrics is as follows:

- H-WMC: A class with more member functions than its peers is more complex and tends to be more fault-prone.
- H-DIT: A class located deeper in a class inheritance lattice is supposed to be more fault-prone because the class inherits a large number of definitions from its ancestors.
- H-NOC: It's usually difficult to test classes with large number of children.

The data collected from the study included the source code of the projects and the data relating to errors found during the testing. GEN++ is a program analyzer used to deduce the CK metrics from the source code. Some additional forms used were Fault report form and Components Origination form. The analysis methodology used is:

- Univariate Logistic Regression- It analyzes the relation between the six CK metrics (WMC, DIT, RFC, NOC, LCOM and CBO) and the probability of fault detection in the classes.

- Multivariate Logistic Regression- It analyzes the metrics on an optimal model rather than predictive model like univariate logistic regression.

The findings of the case study suggest that five (RFC, CBO, LCOM, WMC, DIT) out of six CK Object oriented metrics are practical and useful in predicting the fault proneness during the low level and high level design phase of the software development life cycle. [12]

*B.C An Evaluation of the MOOD set of Object-Oriented Software Metrics*

In this paper MOOD metrics is considered for object oriented design from the viewpoint of measurement theory. Also, their empirical evaluation is considered using three different points. A distinction is made between direct measurement and indirect measurement, internal and external attributes of the process.

In the theoretical validation Encapsulation, Inheritance, Coupling and Polymorphism exist. The below points briefly describes the metrics representing these properties:

- Encapsulation comprises of MHF (Method Hiding Factor) and AHF (Attribute Hiding Factor). MHF and AHF are made to measure the relative amount of hiding, not the quality of information hiding. So, these meet the first four criteria of Kitchenham.
- In Inheritance MIF (Method Inheritance Factor) and AIF (Attribute Inheritance Factor) are used which measure directly the number of inherited methods and attributes respectively. Different programs can have same MIF value and also different MIF value with different amount of inheritance.
- Coupling determines the coupling factor to measure the coupling between classes. There are 2 methods for validating CF. Firstly we can consider CF to be a direct measure of inter-class coupling and Secondly we can consider CF to be an indirect measure of the attributes. It is difficult to pronounce on the validity of CF.
- In Polymorphism the PF (Polymorphism Factor) is used as measure of polymorphism potential. It is an indirect measure of the relative amount of dynamic biding in a system. PF is not a valid metric. But if the metrics discontinuity is removed, it would become valid. [26]

*B.D An Empirical Study of the relationship of Stability Metrics and the QMOOD Quality Models over Software Developed Using Highly Iterative or Agile Software Processes*

The study is conducted to validate the relationship between the Bansiya and Davis total quality index whether it reflects stability over the data sets examined or not. The Quality measures were checked based on the value of three attributes, i.e. SDI (System Design Instability), SDIe (System Design Instability with Entropy) and TQI (Total Quality Index). The agile methods were introduced to check the stability of the program. The agile methods stress on two concepts:

- The unforgiving honesty of the working code.
- The effectiveness of people working together with goodwill.

There were 6 projects inspected which were implemented using extreme programming and out of which 5 projects were student projects and one was highly iterative open source project, where each student project was supposed to have 4 iterations with an exception of project E which due to some reasons only had 3 releases. The developments for all these processes closely follow the twelve principles of agile method as the users have become developers in most open source projects.

The SDI metric was computed for all project iterations by manually inspecting the source code to determine the number of classes whose names have changed, been added, or deleted. The SDIe metric was computed by inspecting the source code for classes added or deleted. The eleven Bansiya and Davis design properties were collected for all iterations. The six quality factors reusability, flexibility, understandability, functionality, extendibility and effectiveness were calculated and based on these values TQI value is calculated, which is the sum of all the quality factors. Within each project we first normalized each measure: SDI, SDIe and TQI by computing the mean and standard deviation for each measure and then subtracted the respective mean and divided by the standard deviation.

The values of SDI, SDIe and TQI are graphed for all iterations of each project and their relationship is noted. The values of SDI, SDIe and TQI depend on the division of the work and the release of code in all iterations. The lack of the division of the work into equal iterations, and the hurried rush to complete the final iteration impacts the TQI value moving it downward in the graph.

The statistical analysis of the code of all the projects concluded that there is a positive correlation between the TQI value and both the stability metrics and the relationship between the SDI and TQI metric is stronger than the relationship between the SDIe and TQI. So further due to the similarity between the SDI and TQI metric, it prompted to use the TQI metric instead of the SDI metric for a stability measure, since the SDI metric requires human participation when analyzing extreme programming projects. [20]

*B.E Measurement of Cohesion and Coupling in OO Analysis Model Based on Crosscutting Concerns*

This paper describes the controlling measurements for Cohesion and Coupling in Object Oriented analysis model based on the notion of crosscutting concerns. It aims to implement the software development quality patterns, Low coupling and high cohesion throughout the software development life cycle. Crosscutting concerns are the parts of a program that rely on or must affect many other parts of the system. Controlling OO analysis model's quality is a crucial aspect of software development as errors introduced in the OO analysis model might propagate throughout the development phases into the final product where to make any correction would require additional efforts and resources.

An OO analysis model is a formal presentation of specifications through which requirements are transmitted to OO analysis model which consists of use cases, class diagrams and domain models.

This paper has introduced measurements inducted at the analysis level to help identify early crosscutting implications in the system. Cohesion measurement is new while the Coupling is an adoption of an existing OO design measure. It addresses the quality patterns in terms of low coupling and high cohesion at requirements analysis stage of software development process when the behavior of the system is modeled as Use cases and the visible part of the system and its relation with real world objects as Domain model.

The main goal is to obtain early feedback at the cohesion and Coupling levels in the analysis model, so as to:

- Reduce Complexity.
- Increase prediction and controlling the scope of changes to a system.
- Designing system with weakest coupled classes thus promoting encapsulation.

**Cohesion Measurement Method:**

Our goal is to measure the cohesion of a use-case (local level) and the cohesion of a use-case model (global level).

$$CL\_UC = |Q| / |P| \qquad 2$$

Q be the set of the similar pairs of scenarios belonging to one use-case and P be the set of all pairs of scenarios belonging to the same use-case.
CL_UC is [0..1], where 1 indicates the highest cohesion and 0 indicates the lack of cohesion in the use-case thus requiring re-analysis of the functional requirements related to this use-case.

$$CL\_UCM = 1-|QM|/|PM| \qquad 3$$

QM is the set of the pairs of similar scenarios and PM is the set of all pairs of scenarios.
CL_UCM is [0...1], where 1 indicates the highest level of cohesion and 0 indicates the lack of cohesion. Higher CL_UCM values indicate that possible cross cuttings are to be identified.

**Coupling Measurement Method:**

MOOD's Coupling Factor measure is used to quantify the existing level of coupling in the domain model.

$$CF = \sum TC_i = 1 \, [\sum TC_j = 1 \, is\_client(C_i, C_j)] / (TC^2 - TC) \qquad 4$$

where,

$$is\_client(C_i, C_j) = \begin{cases} 1, & \text{if } C_c => C_s \wedge C_c \neq C_s \\ 0, & \text{otherwise} \end{cases}$$

The range of the values for the CF is [0...1], where 0 indicates lack of coupling, and 1 is indicates highest possible level of coupling. As a higher value of CF would indicate higher level of coupling between the concepts in the (partial) domain, this value may be considered as an implication of crosscutting requirement(s) to be realized. [3]

*B.F  A Practical Model for Measuring Maintainability*

The research article narrates the number of metrics and its range of values for the measurement of main characteristic of software product quality, which is based on ISO model 9126. Six characteristics of software quality model are functionality, reliability, usability, efficiency, maintainability, and portability. Maintainability is one of the main characteristic of software product quality. ISO 9126 does not facilitate to determine the maintainability based on system's source code. In order to find that, Maintainability index is used to delineate the maintainability of the system, but it has some problem concerning with root-cause analysis, ease of computation, language independence and understandability. The function for MI (Maintainability Index) is:

$$171 - 5.2\ln(HV) - 0.23(CC) - 16.2\ln(LOCPM) + 50.0\sin\sqrt{2.46(CLPM)} \qquad 5$$

Where, HV is Halstead Volume.
  CC is Cyclomatic Complexity.
  LOCPM is Line of Code per Module and
  CLPM is Comment lines per Module

Higher value of MI indicates that system is easy to maintain but the value of MI does not provide the information relating to the characteristic which affect the value of maintainability and what action is required to ameliorate the values which affect the maintainability of the system. The core part of this research paper is to propose such a maintainability model to solve these types of problems. Author has declared such minimum requirements that should be achieved by practical & standard model of maintainability. These requirements are met by some measures, which can be used in model. The characteristics of the measures are as follows:

- The Measure should be independent of technology so it can be applied to system, which has different language and different attributes.
- Every Measure should have clean definition, which is easy to calculate.
- Every Measure should be easy to understand and explain.
- Measure should empower the root cause analysis.

With these requirements, it is easy to formulate alternative maintainability model. Appropriate selection of measure provides the effective connection between source code metrics and ISO 9126 quality characteristics. In the proposed maintainability model system level quality characteristics is mapped into source code measure in two steps:

- Mapping of system characteristics onto source code properties.
- For each Property one or more source code measure are determined.

The properties like volume, Complexity per unit, Duplication, Unit Size and Unit Testing provide the approximation of the main constructive relationships between Code properties and system characteristics. Through the scale it gives ranking to the properties and characteristic of systems. For each of the properties there is well defined range of values through which we can define the value of maintainability. Proposed model provide solution for the problem of Maintainability index. Certain types of modifications are made to the maintainability model according to the different cases.

The Case study provides the standard practical model to measure the maintainability of the system .Every time this model is tested on projects and then modification should be made in order to make the maintainability model effective. [9]

*B.G  Detection Strategies: Metrics-Based Rules for Detecting Design Flaws*

The existence of design flaws in a software system impacts its performance negatively thereby increasing inflexibility and maintenance effort. Software defects symbolize poor quality and bad design practices.  In this section we describe Detection Strategies which provides metrics based rules to determine deviations from good software designs. Many metrics are used to evaluate the performance of software systems but the application of those metrics does not provide enough information regarding the interpretation of the obtained results and the extent to which the metric is accurate. To cater these issues the need for implementing detection strategies was observed.

Detection strategies are constructed using some informal rules depending upon the correlated set of symptoms:

- The first step in the design of detection strategies introduces some rules. The first rule is based on high class complexity. Software complexity is defined as the extent to which a system is difficult to comprehend. The second rule relates to the low class cohesion and the third rule refers to coupling. Cohesion and coupling are important parameters in software design which are inversely related to each other. Good software should have high cohesion and low coupling since cohesion measures the degree to which the methods are related to each other.

- Second step involves selection of appropriate metrics. WMC is used to determine the cumulative complexity of all methods in the class, TCC provides the relative number of directly connected pairs and lastly ATFD represents the count of classes from which a given class accesses attributes.
- The next step is based on adequate filtering mechanism corresponding to each metric. Filtering mechanism deploys data reduction techniques. This process supports the interpretation of the individual metric results. This phase enhances the system's accuracy thereby increasing its correctness. By limiting the methods through constraints it reduces the chances of errors.
- Lastly, these symptoms are correlated using composition operators. In comparison to the filtering technique, composite mechanism is used to support a correlated interpretation of multiple sets of results. It comprises of either logical or set viewpoint to use composite. Simple interpreted language known as SOD is used to automate the detection process using some toolkits. PRODEOOS provides the final results and values of the metrics involved in the detection process.

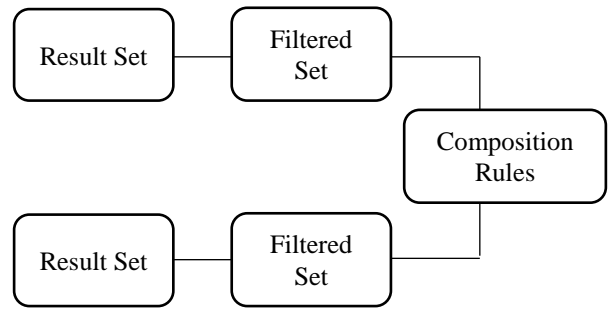

**Figure 5:  Filtration and Composition Mechanism [17]**

Detection strategies should be efficient in properly identifying and localizing the problems in software design fragments and thus helps in achieving high level of abstraction. These conditions can be tested for applicability and accuracy. Experimentation verifies whether these strategies can be defined or not and also, the accuracy of results obtained using either automatic or manual approach. Based on the experimentation nearly 10 design flaw categories were examined for accuracy rate and it was observed that the suspected entity was actually affected by the design flaw. These results reveal the effectiveness of the approach. Thus, application of detection strategies makes the system more flexible and understandable and such a system is easy to maintain and modify.

Although, the mechanism provides several possibilities for determining design flaws but still there exists a need to uncover more techniques in the environments where detection strategies cannot be established. [17]

*B.H  Voronoi Treemaps for the Visualization of Software Metrics*

The paper was proposed to show how hierarchy based visualization approach is applied for software metrics using Treemaps. Author has introduced Voronoi Treemaps; a polygon based layouts which focus more on aspect ratio between width and height of the objects. It also identifies boundaries between and within the hierarchy levels in the Treemap. There are many approaches are available in software visualization but only few of them shows proper visualization metric. In some methods like Slice-and-Dice Treemaps and Squarified Treemaps has aspect ratio and hierarchy level problem respectively.

Voronoi Treemaps has some optimization criteria for the shape of treemap object. It follows below criteria:

- Distribution object must utilize the given area.
- Object should distinguish themselves.
- Object should be compact, it means ratio of objects height and width should coverage to one.

Below are steps to calculate Voronoi Treemap:

- The Centroidal Voronoi Tessellation (CVT) is computed by determining the Voronoi tessellation of a given point set.
- With basic CVT we are able to compute Treemap layouts in which every leaf node in the hierarchy has the same value, but our goal is to find different value. So we have to recalculate CVT using this formula:

distance(c;q)Circle := distance(p;q) – r           6

- In last step, adjustments of the radii in every iteration step, the computation will end up in a stable state. CVT has been computed and Voronoi Polygon is extracted.

The result of case study suggests new method for computing layouts of Treemaps, which is based on polygons.

Advantages of these visualization methods are:

- Good aspect ratio, interactive zooming and transparency between components.
- Visualizes the outbound calls of classes by other classes of the software system by 'ArgoUML'.
- Visualizes the lines of code (LOC) of all files of the software system by 'JFree'. [5]

*B.I  Summary*

Design metrics play a crucial role in determining the quality of the system. In this section we describe, compare and observe the impact of the metrics discussed in the previous section on the system's design and analyze how it affects the non-functional characteristics.

MARF provides a framework for measuring pattern recognition algorithms. Its applications when implemented enhance system's extensibility, modifiability, maintainability, adaptability and efficiency. Almost similar features are captured by GIPSY with an exception to high degree of flexibility which is achieved through language independent approach. Cumulatively these attributes strengthen the understandability and functionality of the system.

Based on above features we can measure the performance of each metric and determine the quality attributes affected by them. We divide the study into two parts:

**PART A:** Metrics where code is not analyzed

**QMOOD:**

QMOOD metrics are used to determine high level quality characteristics of object oriented design. We can assess the quality of the system without making use of source code.

Following table gives summary of QMOOD metrics and their impact on quality characteristics covered in the above section:

Table 7: Metrics Summary (QMOOD)[35]

| Metric | Design Property | Quality Attributes Affected |
|---|---|---|
| DAM | Encapsulation | Understandability |
| CAMC | Cohesion | Extendibility, Effectiveness and Reusability |
| ANA | Abstraction | Effectiveness, Understandability and Extendibility |
| MFA | Inheritance | Extendibility, Effectiveness and Reusability |
| NOP | Polymorphism | Extendibility, Flexibility, Functionality, Effectiveness, Understandability and Reusability |
| DSC | Design Size | Functionality, Understandability and Reusability |
| CIS | Messaging | Functionality and Reusability |

**PART B:** Metrics where code is analyzed

MOOD and CK metrics comes under this category.

**CK Metrics:**

Table 8: Metrics Summary (CK)[35]

| Metric | Design Property | Quality Attributes Affected |
|---|---|---|
| WMC | Complexity | Understandability |
| DIT | Inheritance | Extendibility, Effectiveness and Reusability |
| NOC | Hierarchy | Functionality |
| CBO | Coupling | Understandability, Flexibility, Extendibility and Reusability |
| RFC | Polymorphism | Extendibility, Flexibility, Functionality, Effectiveness, Understandability and Reusability |

**MOOD:**

MOOD metrics are used provide the quality related summary of an object oriented project. The design properties captured by MOOD are:

- Encapsulation
- Coupling
- Inheritance, and
- Polymorphism

Following table gives summary of MOOD metrics and their impact on quality characteristics:

Table 9: Metrics Summary (MOOD)[35]

| Metric | Design Property | Quality Attributes Affected |
|---|---|---|
| MHF / AHF | Encapsulation | Understandability, Flexibility and Effectiveness |
| MIF / AIF | Inheritance | Extendibility, Effectiveness and Reusability |
| PF | Polymorphism | Extendibility, Flexibility, Functionality, Effectiveness, Understandability and Reusability |
| CF | Coupling | Understandability, Flexibility, Extendibility and Reusability |

Below figure gives the relation between these design properties and quality attributes. It shows whether the design property has direct or inverse relation with the attributes.

| | Reusability | Flexibility | Understandability |
|---|---|---|---|
| **Design Size** | ↑ | | |
| **Hierarchies** | | | |
| **Abstraction** | | | |
| **Encapsulation** | | ↑ | ↑ |
| **Coupling** | | | |
| **Cohesion** | ↑ | | ↑ |
| **Composition** | | ↑ | |
| **Inheritance** | | | |
| **Polymorphism** | | ↑ | |
| **Messaging** | ↑ | | |
| **Complexity** | | | ↓ |

| | Functionality | Extendibility | Effectiveness |
|---|---|---|---|
| **Design Size** | ↑ | | |
| **Hierarchies** | ↑ | | |
| **Abstraction** | | ↑ | ↑ |
| **Encapsulation** | | | ↑ |
| **Coupling** | | | |
| **Cohesion** | ↑ | | |
| **Composition** | | | ↑ |
| **Inheritance** | | ↑ | ↑ |
| **Polymorphism** | ↑ | ↑ | ↑ |
| **Messaging** | ↑ | | |
| **Complexity** | | | |

**Figure 6: Relation between attributes and properties [23]**

In order to select the most appropriate metrics we formulate some ground rules:

- Metric selected should be able to identify goals clearly.
- It should be measurable.
- It should be capable of deriving some hypothesis regarding the key quality factors.
- It should collect data both in quantitative and qualitative terms.
- It should provide a way for improvement.

Thus based on above rules we select the most appropriate metrics as per the system design. Prioritization is done considering 3 cases.

**Case 1: Prioritization at Metrics level**

CK and QMOOD metrics contains similar components and they produce the components which are statistical in nature which are efficient in detecting the classes which are vulnerable to errors and design flaws as compared to the MOOD metrics.[24]

Thus we finalize the metrics as:

QMOOD / CK > MOOD

**Case 2: Prioritization based on Design Parameters**

Every software system is intended to achieve Maintainability, Flexibility, and Extendibility. In order to obtain above features, encapsulation is highly important in the design. On the other hand Inheritance increases Extendibility as the class is able to access the methods and attributes of another class but decreases fault detection a class located deeper in a class inheritance lattice is supposed to be more fault-prone because the class inherits a large number of definitions from its ancestors. Poly means existing in many forms. Polymorphism feature in system design enables flexibility. Hence we are prioritizing our metrics based on below mentioned relation of Design properties:

Encapsulation > Inheritance / Polymorphism > Design Size

| Metric Method | Design Property |
|---|---|
| DAM, MHF/AHF | Encapsulation |
| ANA, MFA, NOP, DIT, RFC, MIF/AIF, PF | Inheritance / Polymorphism |
| DSC | Design Size |

Thus, we finalize the metrics as:

DAM, MHF/AHF > ANA, MFA, NOP, DIT, RFC, MIF/AIF, PF > DSC

**Case 3: Prioritization based on Quality Attributes**

Software should be understandable first. Software which is well understandable makes it more maintainable and testable. On the basis of analysis based on MARF and GIPSY we prioritize these two traits. High cohesion within program modules makes it readable, maintainable and reusable as classes with core functions are not corrupted with useless functions. We will be using DAM and CAM metric to evaluate this.[33][34]

Extendibility of software can be measured through the presence of level of abstraction, inheritance and polymorphism. Abstraction promotes inheritance and polymorphism. A high level of Abstraction and inheritance of software depicts that it is reusable and extendible but at the same time if a class has huge number of child classes it makes the program more complex which reduces understandability and maintainability of the software.

Design size of software elaborates the number of classes existing in the software. Cohesion shows relatedness between methods by assessing method parameters and attribute types and messaging relates to the concept of coupling which shows the use of methods of one class in other classes. A high number of classes or low cohesion amongst methods or high coupling or message passing between classes shows that the software is complex and is difficult to maintain and understand. Since systems are made for humans and for their better functioning and improvements they require human intervention. Hence everything comes down to Understandability of the designed software system. Hence we are prioritizing our metrics based on below mentioned relation of quality attributes:

Understandability > Extendibility > Reusability

| Metric Method | Quality Parameter |
|---|---|
| DAM, CAMC | Understandability |
| ANA, MFA, NOP | Extendibility |
| DSC, CIS | Reusability |

Thus, on the basis of above findings we finalize the metrics as:

DAM, CAMC > ANA, MFA, NOP > DSC, CIS

Out of all the 3 cases we analyzed, our major concentration will be on the metrics category based on Quality Attributes.

*C. Methodology*

*C.A Metrics with Tools: Logiscope and McCabe*

*Logiscope*

We have used Kalimetrix Logiscope to verify the source code for MARF and GIPSY. Logiscope is a set of software analyzers that determines the quality of software. The quality of code is measured in terms of software metrics, which indicates its complexity, testability, and understandability along with descriptive statistics. The main advantage of using Logiscope is that it helps us to deliver great quality software products with increased software maintainability and reliability.

In order to determine the maintainability of the code, a quality report is generated to study different aspects involved in its evaluation.

Maintainability is defined as the ease of maintaining a software system. It is a part of software maintenance activity. Maintainability factor is used to determine the maintainability of a program.

As per ISO/IEC 9126-1:2001 Maintainability Factor is defined as the capability of a software product to be modified.

Modifications may refer to:

- Corrections
- Improvements
- Adaptation of the software to changes in environment,
- Or adapting with respect to the changing in requirements and functional specifications.

The extent of analyzability, changeability, testability and stability of a system directly impacts its maintainability.

The formula used to determine the values of maintainability factor is:

$$\text{MAINTAINABILITY} = \text{ANALYZABILITY} + \text{CHANGEABILITY} + \text{STABILITY} + \text{TESTABILITY} \qquad 7$$

Thus we observe the use of four criteria's to compute its value. Below section defines the ISO/IEC 9126-1:2001 description of these criteria's:

- **ANALYZABILITY** - It is defined as the capability of a software product to be diagnosed for deficiencies or causes of failures in the software, or for the parts to be modified and to be identified.

Formula for calculating its value is:

$$\text{ANALYZABILITY} = cl\_wmc + cl\_comf + in\_bases + cu\_cdused \qquad 8$$

Thus we may say that analyzability is a direct measure of system's complexity, comment rate, count of the classes used directly and the number of base classes.

- **CHANGEABILITY** - The capability of the software product to enable a specified modification to be implemented.

Formula for calculating its value is:

$$\text{CHANGEABILITY} = cl\_stat + cl\_func + cl\_data \qquad 9$$

Thus we conclude that changeability is a direct measure of system's total number of methods, statements and attributes.

- **STABILITY** - The capability of the software product to avoid unexpected effects from modifications of the software.

Formula for calculating its value is:

$$\text{STABILITY} = cl\_data\_publ + cu\_cdusers + in\_noc + cl\_func\_publ \qquad 10$$

Thus we may say that stability is a direct measure of the sum of system's public attributes, direct user classes, number of children and count of the number of public methods.

- **TESTABILITY** – It is the capability of the software product to enable modified software to be validated

Formula for calculating its value is:

$$\text{TESTABILITY} = cl\_wmc + cl\_func + cu\_cdused \qquad 11$$

Following table gives detailed description of code characteristics measured by the above criteria's and their influence on the system if their values are high or low:

Table 10: Code Characteristics and Value analysis [30]

| Metric name | Code Characteristic | Value Analysis |
|---|---|---|
| cl_wmc | It denotes weighted methods per Class which is the sum of the static complexities of the class methods. Static complexity is represented by the cyclomatic number of the functions.<br><br>cl_wmc = SUM (ct_vg) | Code with higher value of weighted method per class makes it more complex. Hence it is difficult to maintain and understand the code. Higher the value of WMC lower is the analyzability. Therefore, M1 is better value than M2. |
| cl_comf | It represents the class comment rate which is the ratio between the number of lines of comments in the module and the total number of lines:<br><br>cl_comf = cl_comm / cl_line<br><br>where cl_comm – Number of comment lines and,<br>cl_line – Total number of lines. | Code with greater comment rate makes easier to understand and analyze. It is easier to make changes, comprehend and maintain the code. Less effort would be required for testing. Hence M2 is better than M1. |
| cu_cdused | It is defined as the number of classes used directly by the current class. | With more number of direct used classes less efforts are required to test. Also, it is easy to understand the code to make changes because of the higher value. Code is more stable. Hence M2 is better than M1. |
| in_bases | It gives a measure of the number of classes from which the class inherits directly or not. If multiple inheritance is not used the value of in_bases is equal to the value of in_depth | Higher value M2 is better when compared to lower value M1 since the lesser the value the lesser is the complexity due to multiple inheritance. |
| cl_stat | It denotes number of executable statements in all methods and initialization code of a class. | Higher value takes more efforts to analyze. Also, more efforts are require for testing or to make changes. Hence M1 is better than M2. |
| cl_func | It indicates the total number of methods declared inside the class declaration. | Analyzing the code with more methods for faults or failures is difficult. Also, it is easier to modify and test code with fewer methods as the code with less number of methods is more stable. Hence M1 is better than M2. |
| cl_data | It represents the total number of attributes declared inside the class declaration | Lesser the number of attributes more easy it is to maintain, modify, correct and test the code. Code with less attributes is more stable as the value of one attribute may depend or would have been derived from another attribute being refactored. Hence M1 is better than M2. |
| cl_data_publ | It is defined as the number of attributes declared in the public section or in the public interface of a Java class. | Higher the value of public attributes, the more likely class provides access to its attributes, lesser the efforts to modify higher changeability. |

| | | | | | | |
|---|---|---|---|---|---|---|
| | | Lesser the efforts to understand and test. And it is more stable code. Hence M2 is better than M1. | | | | makes code more complex to maintain, understand, modify and test. Hence M1 is better than M2. |
| cu_cdusers | It refers to the number of classes which use directly the current class. | With higher value, higher the use of current class. Code is more stable. Easy to understand the code. Changeability is higher. Hence M2 is better than M1 | | | | |
| in_noc | Number of classes which directly inherit from the current class gives the value of number of children. | Lesser the value more easy it becomes to analyze. If the value is high, modifications induce changes thus reducing the stability and require more efforts to test and validate. Hence M1 is better than M2. | | | | |
| cl_func_publ | Number of methods declared in the public section is denoted by this metric. | Analyzing the code with more number of public methods for fault or failures is easier. Also, it is more likely methods provide access to modify and test cases will be less. Hence M2 is better than M1. | | | | |
| Cl_comm | Number of lines of comment | Code with more number of comments make easier to understand and helps to perform maintaining operations. It is also better to distribute the comments evenly rather than placing them at the beginning. Hence M2 is better than M1. | | | | |
| Cl_line | Number of lines | With higher value it | | | | |

In the subsequent sections we process the measurement feedback loop for the implementation of MARF and GIPSY. Measurement feedback loop consists of 3 stages:

- Extract: In this phase we have extracted the information pertaining to Maintainability at the class factor, criteria and metric level.
- Evaluate: In this phase we evaluate and analyze the obtained results.
- Execute: Finally, we execute the findings to provide recommendations.

**Phase 1: EXTRACT**

**MARF**

On the basis of the source code we have determined the value of maintainability at 3 levels i.e. at factor level, criteria level and at metric level.

**A.** Class Factor Level

The pie chart represents system's maintainability at Class Factor Level.

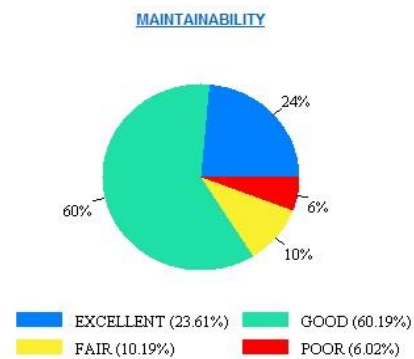

**Figure 7: Maintainability factor MARF**

**B.** Class Criteria Level

On a class criteria level the pie charts are generated for four quality criteria's:

- Analyzability

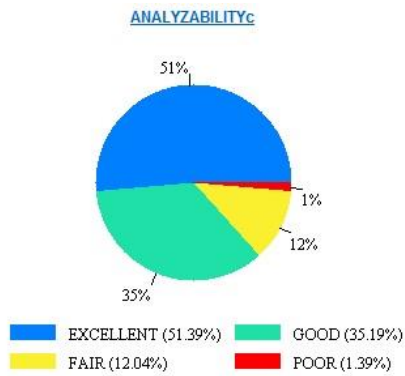

Figure 8: Analyzability MARF

- Changeability

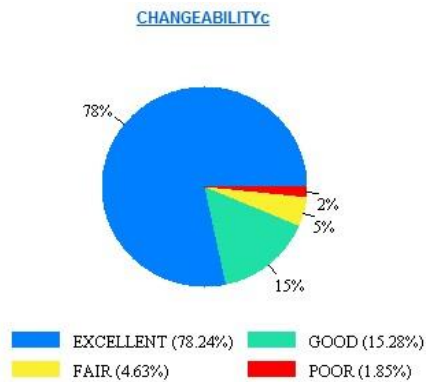

Figure 9: Changeability MARF

- Stability

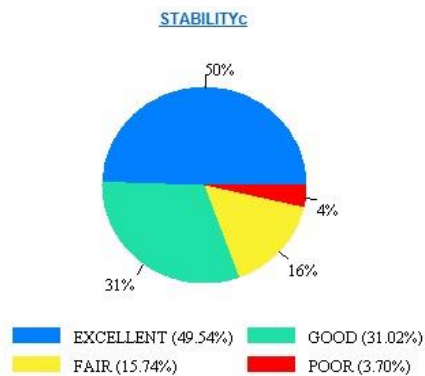

Figure 10: Stability MARF

- Testability

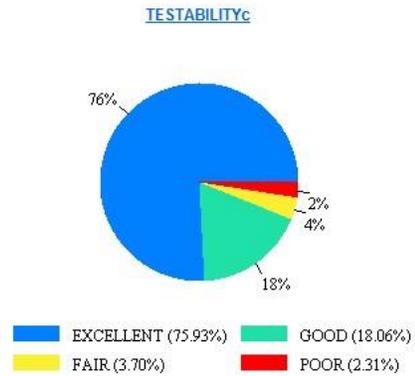

Figure 11: Testability MARF

**C.** Class Metric Level

At metric level we have chosen two classes namely marf.util.Matrix and test to demonstrate the metrics.

The first class belongs to the excellent category while the second class belongs to the poor category.

Kiviat plots are generated for both the classes to make comparisons in the metric values.

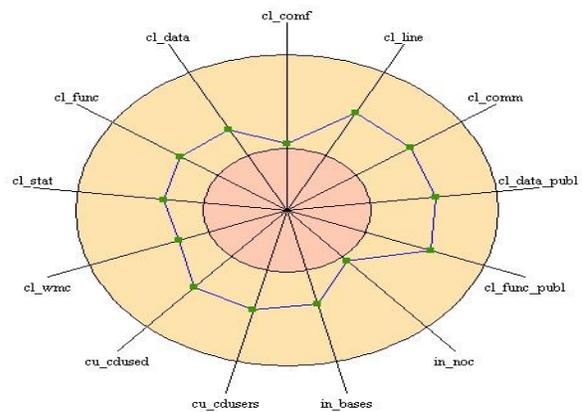

Figure 12: Kiviat Graph marf.util.Matrix

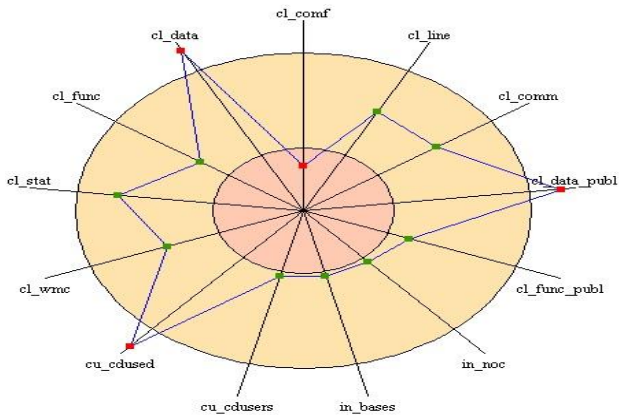

Figure 13: Kiviat Graph test

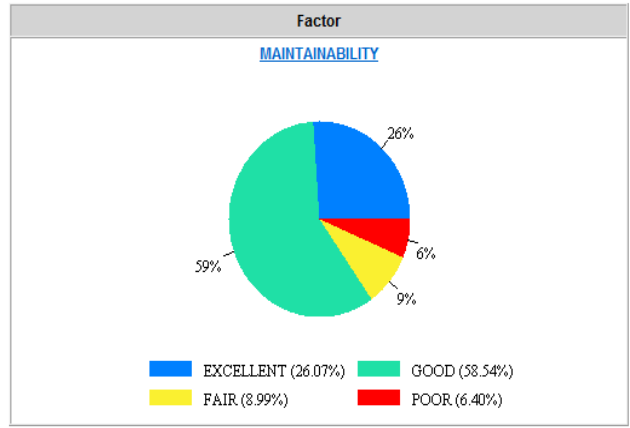

Figure 14: Maintainability factor GIPSY

## GIPSY

Maintainability of entire GIPSY application is represented pictorially using pie charts at 3 levels i.e. Factor level, Criteria level and Metric level.

**A.** Class Factor Level

The pie chart presents system's maintainability at Class factor level.

**B.** Class Criteria Level

On a class criteria level the pie charts are generated for four quality criteria's:

- Analyzability

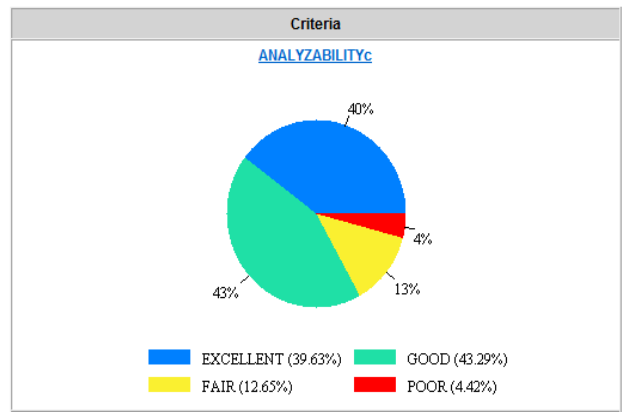

Figure 15: Analyzability GIPSY

- Changeability

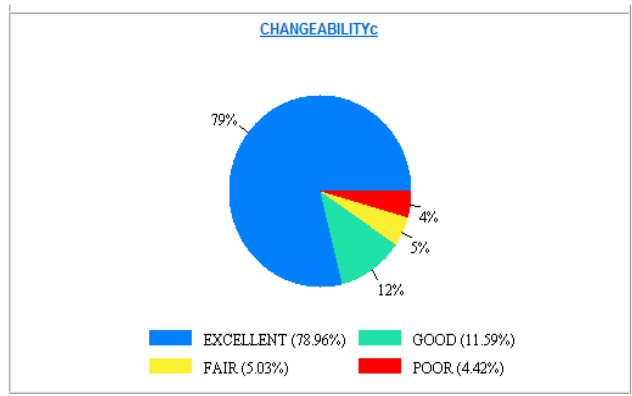

Figure 16: Changeability GIPSY

- Stability

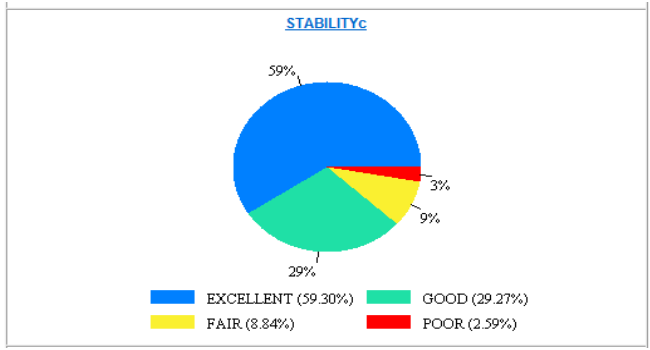

Figure 17: Stability GIPSY

- Testability

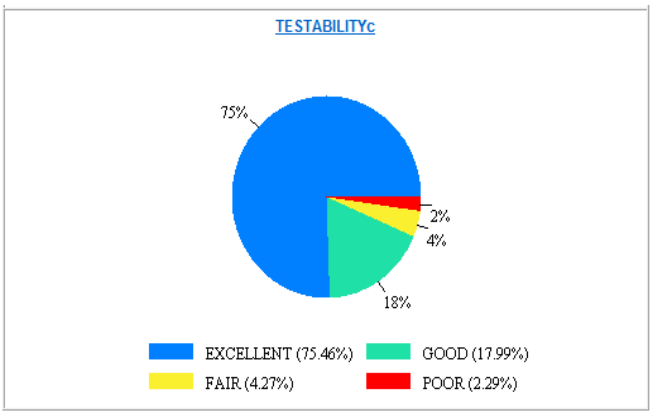

Figure 18: Testability GIPSY

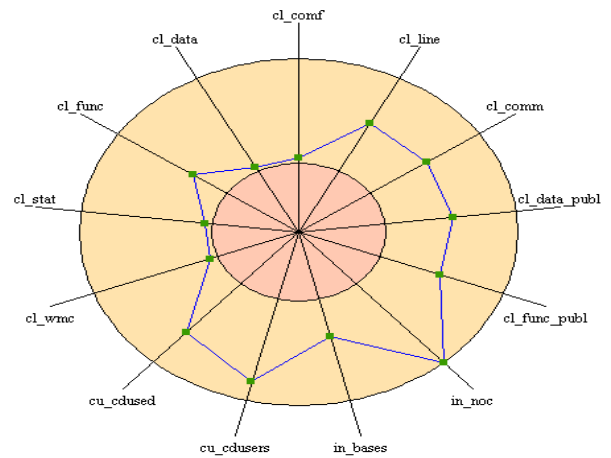

Figure 19: Kiviat Graph GIPC.ICompiler

**C. Class Metric Level**

At metric level we have chosen two classes namely gipsy.GIPC.ICompiler and gipsy.GEE.GEE to demonstrate the metrics.

The first class belongs to the excellent category while the second class belongs to the fair category.

Kiviat plots are generated for both the classes to make comparisons in the metric values.

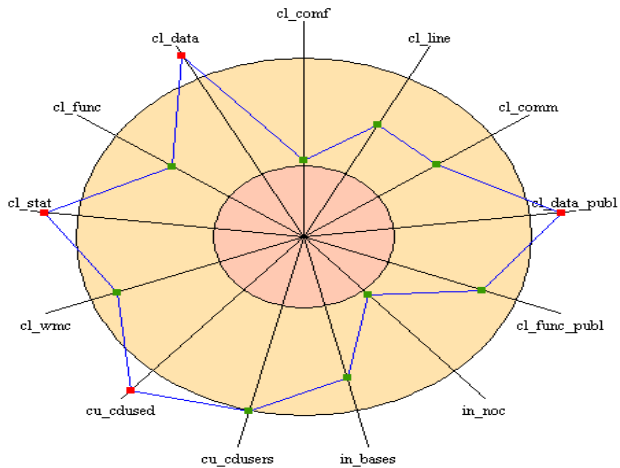

Figure 20: Kiviat Graph gipsy.GEE.GEE

**Phase 2: EVALUATE**

**Comparative Analysis: MARF & GIPSY**

The maintainability and its corresponding criteria curve discussed in above section for MARF and GIPSY generated via Logiscope provide data in terms of excellent, good, fair and poor code quality.

Table 11: Code Quality Level Percentage MARF

| Levels | MARF | | | | |
| --- | --- | --- | --- | --- | --- |
| | Maintainability | Analyzability | Changeability | Stability | Testability |
| Excellent | 24 | 51 | 78 | 50 | 76 |
| Good | 60 | 35 | 15 | 31 | 18 |
| Fair | 10 | 12 | 5 | 16 | 4 |
| Poor | 6 | 1 | 2 | 4 | 2 |

Table 12: Code Quality Level Percentage GIPSY

| Levels | GIPSY | | | | |
| --- | --- | --- | --- | --- | --- |
| | Maintainability | Analyzability | Changeability | Stability | Testability |
| Excellent | 26 | 40 | 79 | 59 | 75 |
| Good | 59 | 43 | 12 | 29 | 18 |
| Fair | 9 | 13 | 5 | 9 | 4 |
| Poor | 6 | 4 | 4 | 3 | 2 |

This data is used and a double bar chart is plotted to perform comparative analysis at Factor and Criteria Level.

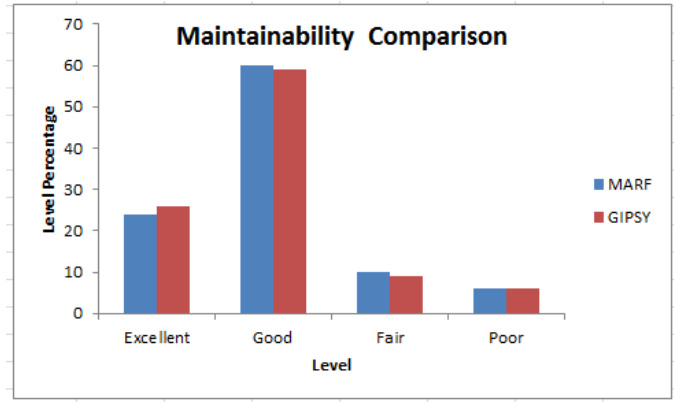

Figure 21: Maintainability Comparison MARF & GIPSY

Maintainability is an essential parameter for determining the quality of a software system. The above bar graph presents a differentiation in the maintainability % level for the two applications in consideration namely MARF and GIPSY. Based on these levels we rank the quality of the code.

Worst quality code comprises of the fair and poor category classes. MARF consists of in total 16% of bad quality code in comparison to GIPSY which has 15% of bad quality code. The 1% difference lies because of the existence of poor classes in the fair category.

From analyzability perspective, 13% of the code in MARF and 17% of the code in GIPSY falls under bad category.

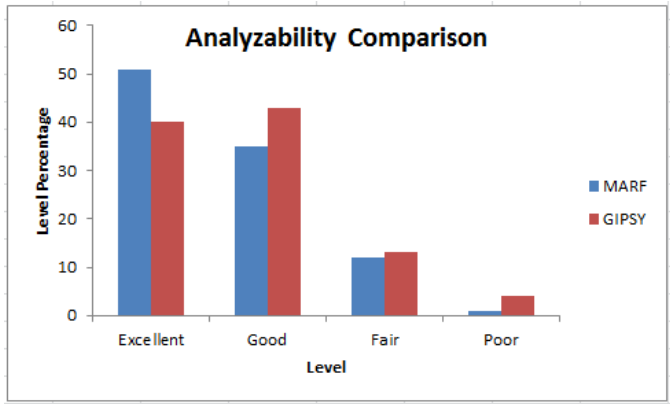

Figure 22: Analyzability Comparison MARF & GIPSY

From changeability perspective, 7% of the code in MARF and 9% of the code in GIPSY falls under bad category.

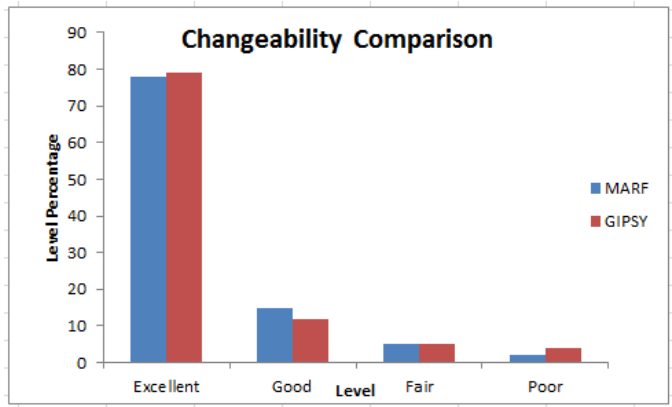

Figure 23: Changeability Comparison MARF & GIPSY

From stability perspective, 20% of the code in MARF and 12% of the code in GIPSY falls under bad category.

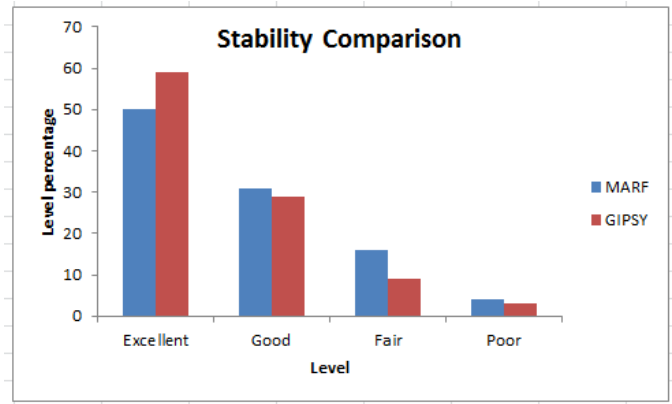

Figure 24: Stability Comparison MARF & GIPSY

From testability perspective, 6% of the code in MARF and 6% of the code in GIPSY falls under bad category.

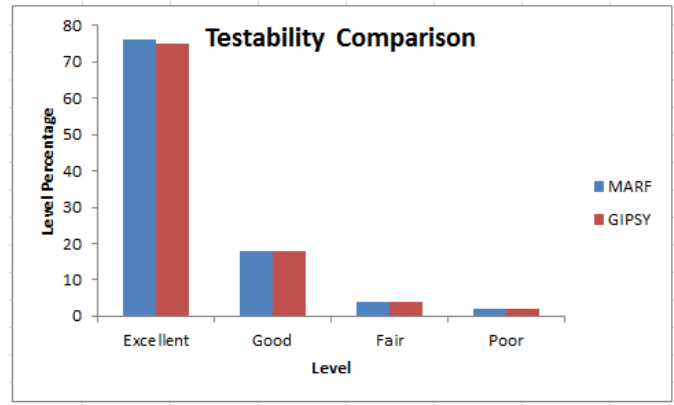

Figure 25: Testability Comparison MARF & GIPSY

Using these results we generate a ranking matrix for MARF and GIPSY. The presence of a tick indicates a higher ranking in the corresponding Factor/Criteria.

Table 13: Ranking Matrix

| Class Level | Quality Attribute | MARF | GIPSY |
|---|---|---|---|
| Factor | Maintainability |  | ✓ |
| Criteria | Analyzability | ✓ |  |
| Criteria | Changeability | ✓ |  |
| Criteria | Stability |  | ✓ |
| Criteria | Testability | ✓ | ✓ |

Fair and poor section corresponds to the worst quality code. Classes included in these sections are specified in following figures. We make use of this data to list the classes present in this section for MARF and GIPSY.

## MARF

## GIPSY

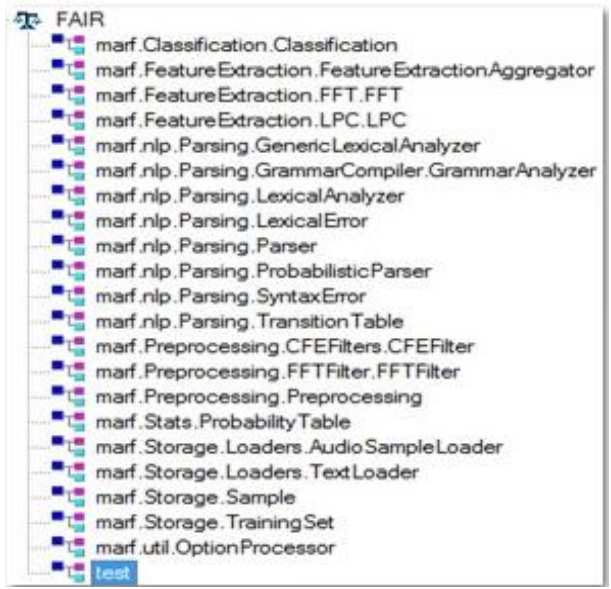

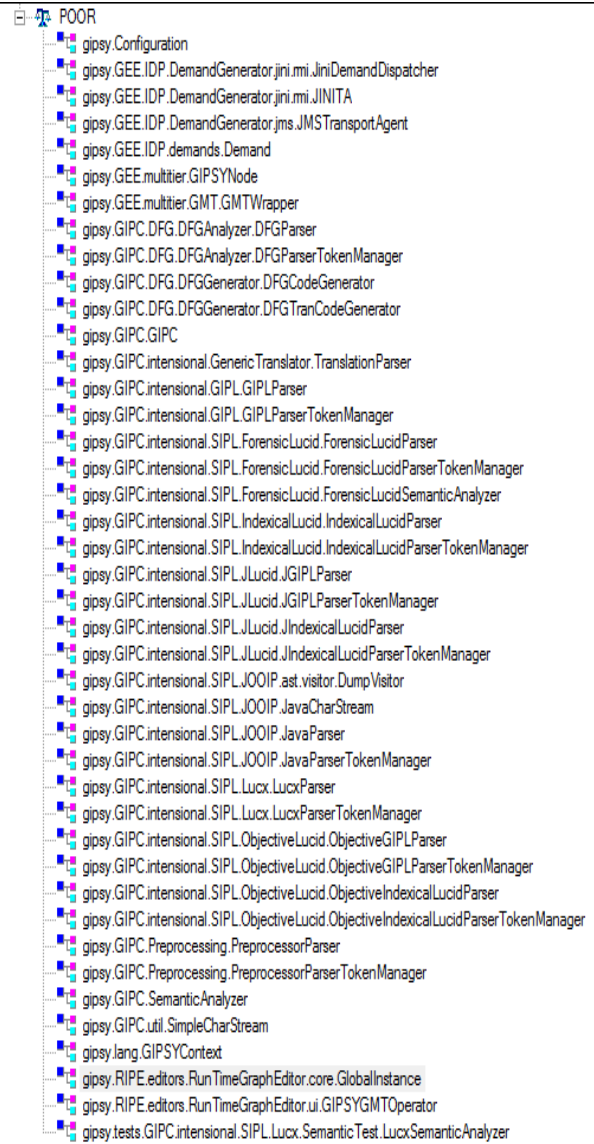

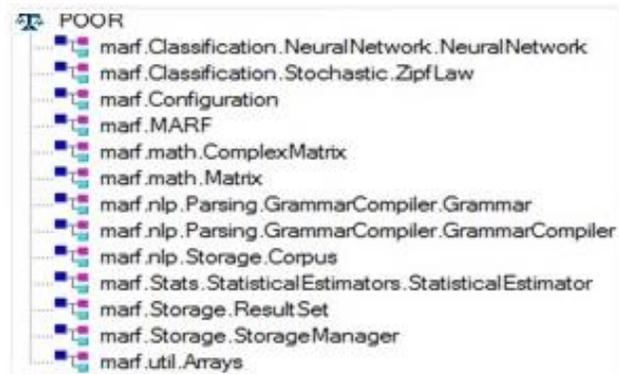

Figure 26: List of Fair & Poor classes MARF

Figure 27: List of Fair & Poor classes GIPSY

From the above listed classes in the worse code quality section we determine two classes which are most ineffective in terms of their quality, performance and consistency. We divide this analysis in two parts and observe the results:

**Case 1:** Intra- MARF Comparison

In this attempt we compare two worst quality classes within the MARF application from the worst quality code section and visualize the same through a Kiviat graph.

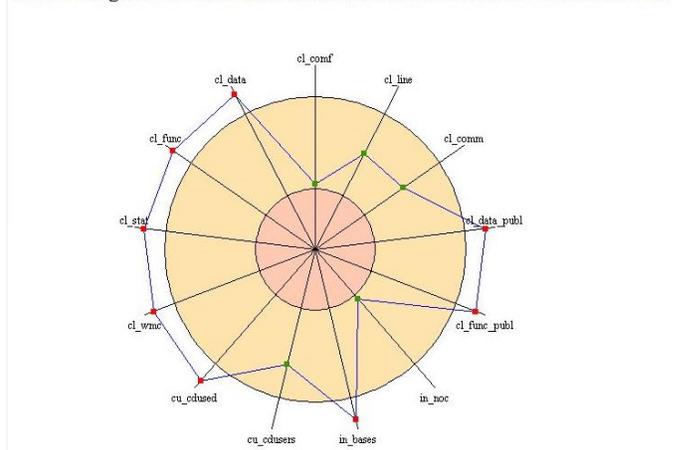

Figure 28: Kiviat Graph Neural Network Class

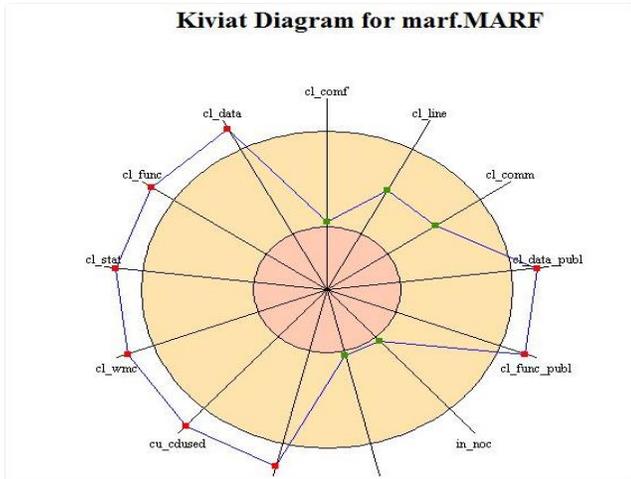

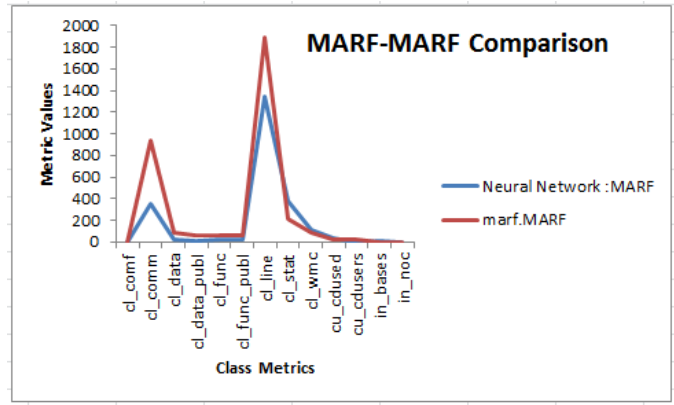

Figure 30: MARF Classes Comparison

Based on Kiviat analysis we conclude that both the classes are poor in Maintainability. Criteria level bifurcation is listed below:

| Factor/Criteria | Neural network | marf.MARF |
|---|---|---|
| Maintainability | Poor | Poor |
| Analyzability | Poor | Fair |
| Changeability | Poor | Poor |
| Stability | Fair | Poor |
| Testability | Poor | Poor |

**Case 2:** Intra- GIPSY Comparison

In this attempt we compare two worst quality classes within the GIPSY application from the worst quality code section.

| Metric : marf.MARF | Value | Min | Max |
|---|---|---|---|
| cl_comf: Class comment rate | 0.49 | 0.20 | +∞ |
| cl_comm: Number of lines of comment | 932 | -∞ | +∞ |
| cl_data: Total number of attributes | 84 | 0 | 7 |
| cl_data_publ: Number of public attributes | 64 | 0 | 0 |
| cl_func: Total number of methods | 58 | 0 | 25 |
| cl_func_publ: Number of public methods | 55 | 0 | 15 |
| cl_line: Number of lines | 1892 | -∞ | +∞ |
| cl_stat: Number of statements | 219 | 0 | 100 |
| cl_wmc: Weighted Methods per Class | 82 | 0 | 60 |
| cu_cdused: Number of direct used classes | 26 | 0 | 10 |
| cu_cdusers: Number of direct users classes | 20 | 0 | 5 |
| in_bases: Number of base classes | 0 | 0 | 3 |
| in_noc: Number of children | 0 | 0 | 3 |

Figure 29: Kiviat Graph Marf Class

On reviewing the above two classes, we notice that both the classes have 8 out of control points which stretch beyond the respective upper control limits but with a distinction of one parameter i.e. cl_cdusers ( Neural Network class ) and in_bases ( MARF class ).

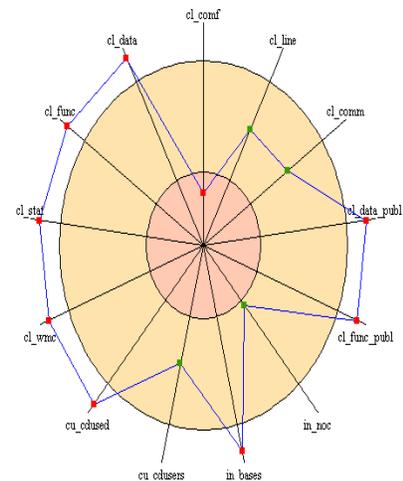

Table 14: MARF Metric Values

| Metric | Neural Network :MARF | marf.MARF |
|---|---|---|
| cl_comf | 0.26 | 0.49 |
| cl_comm | 354 | 932 |
| cl_data | 17 | 84 |
| cl_data_publ | 8 | 64 |
| cl_func | 27 | 58 |
| cl_func_publ | 21 | 55 |
| cl_line | 1348 | 1892 |
| cl_stat | 372 | 219 |
| cl_wmc | 115 | 82 |
| cu_cdused | 33 | 26 |
| cu_cdusers | 3 | 20 |
| in_bases | 6 | 0 |
| in_noc | 0 | 0 |

Figure 31: Kiviat Graph GIPSYGMTOperator Class

On visualizing the Kiviat graph for classes gipsy.RIPE.editors.RunTimeGraphEditor.ui.GIPSYGMTOperator and gipsy.GIPC.util.SimplecharStream, we observe that the first class has 9 ouliers, 8 out of which lie beyond the maximum value and 1 below the minimum value. Metrics representing these points are cl_data, cl_func, cl_stat, cl_comf, cl_wmc, cl_cdused, cl_data_publ, cl_func_publ, in_bases. The impact of the high and low values of these metrics is already discussed in Table 8.

The out of control points indicate the issues in the class. The latter class has 8 out of control points namely cl_comf, cl_data, cl_data_publ, cl_func_publ, cl_func, cl_stat, cl_wmc, cl_cdusers.

**Table 15: GIPSY Metric Values**

| Metric | SimpleCharStream:GIPSY | GIPSYGMTOperator:GIPSY |
|---|---|---|
| cl_comf | 0.1 | 0.19 |
| cl_comm | 45 | 160 |
| cl_data | 17 | 44 |
| cl_data_publ | 2 | 8 |
| cl_func | 38 | 37 |
| cl_func_publ | 33 | 19 |
| cl_line | 464 | 859 |
| cl_stat | 202 | 311 |
| cl_wmc | 67 | 69 |
| cu_cdused | 9 | 51 |
| cu_cdusers | 20 | 2 |
| in_bases | 0 | 6 |
| in_noc | 0 | 0 |

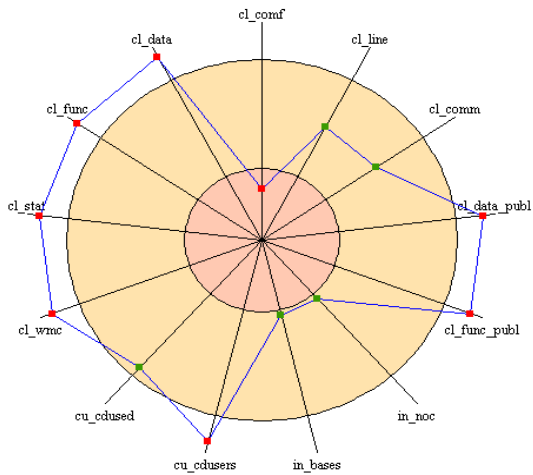

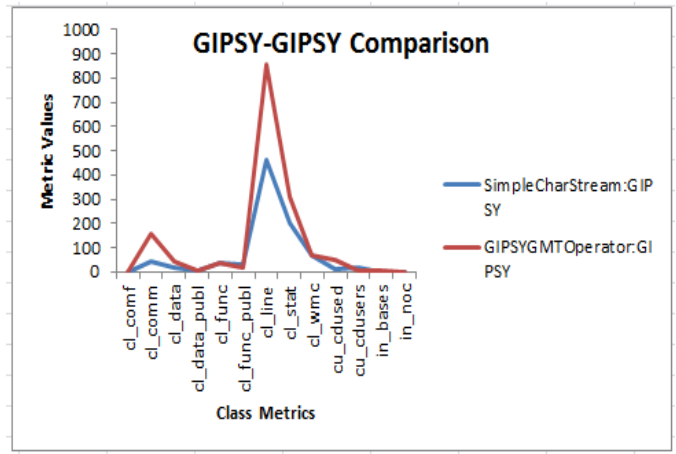

Figure 33: GIPSY-GIPSY Classes Comparison

Figure 32: Kiviat Graph Simple Char Stream Class

Kiviat is a multi-vector line graph which represents different metrics with their values and the ideal range.

The GIPSYGMTOperator class is poor when it comes to Analyzability, Changeability and testability but fair in terms of overall stability while the SimplecharStream class is poor in Changeability and Stability whereas it has a fair quality in Analyzability and Testability. Thus, at metric level the latter class is better as compared to the first class. Analyzability is defined as the property if a software system to be easily understood. It acts as an important prerequisite for editing and changing purposes. If a class is not easy to comprehend and change, then it becomes difficult to test and improve further.

| Factor/Criteria | GIPSYGMT Operator | Simplechar Stream |
|---|---|---|
| Maintainability | Poor | Poor |
| Analyzability | Poor | Fair |
| Changeability | Poor | Poor |
| Stability | Fair | Poor |
| Testability | Poor | Fair |

**Case 3:** Inter-application Comparison (MARF – GIPSY)

In this attempt we compare one worse quality class from MARF and GIPSY each and analyze the results.

Therefore, we compare Neural Network from MARF and GIPSYGMTOperator class from GIPSY based on the previous filteration from worst quality code.

**Table 16: MARF-GIPSY Metric Values**

| Metric | Neural Network :MARF | GIPSYGMTOperator:GIPSY |
|---|---|---|
| cl_comf | 0.26 | 0.19 |
| cl_comm | 354 | 160 |
| cl_data | 17 | 44 |
| cl_data_publ | 8 | 8 |
| cl_func | 27 | 37 |
| cl_func_publ | 21 | 19 |
| cl_line | 1348 | 859 |
| cl_stat | 372 | 311 |
| cl_wmc | 115 | 69 |
| cu_cdused | 33 | 51 |
| cu_cdusers | 3 | 2 |
| in_bases | 6 | 6 |
| in_noc | 0 | 0 |

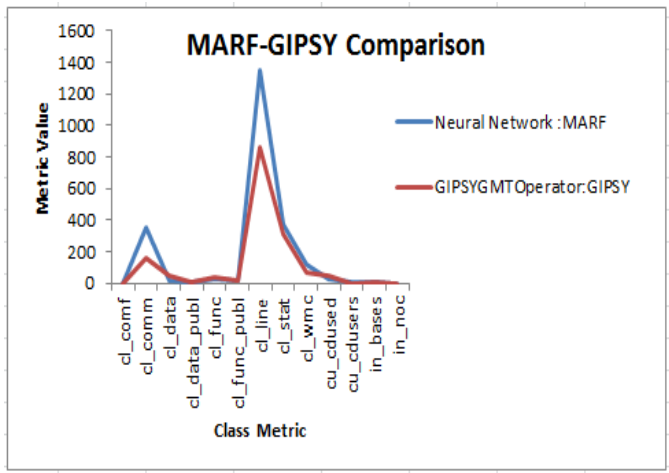

**Figure 34: MARF-GIPSY Classes Comparison**

The line chart elaborates the metric values comparison between the worst among worst classes of both MARF and GIPSY. By critical examination of all the values and cases we conclude that Neural Network is the worst class when compared with the worst class of GIPSY.

**Phase 3: EXECUTE**

In this section we suggest recommendations to improve the quality of the code:

**A.** At Class Factor/Criteria Level

Recommendations for identified MARF Classes:

- Efforts should be made to reduce the coupling between the classes to which it is highly associated. It will make it more maintainable and changeable. Such a design would be simpler to understand.[32]
- Efforts should be made to reduce coupling between the native class package and other packages which would increase the stability factor and enhance changeability if required in future.
- Debugging code should be removed as it can cause trouble during production and can sometimes be a critical issue during maintenance. Hence it should be removed. It can reduce testability but it can be ignored if we look from broader perspective.
- Certain coding standards should be followed which could make the code easy to understand, analyse, maintain and change (For e.g. The ones shown in summary section)
- A blank constructor declared as private is a mischievous code which can cause trouble during maintenance phase or in case of future changes. Hence it should be avoided.

Recommendations for identified GIPSY Classes:

- There were empty declarations found for the certain methods which can be removed as they are increasing code complexity. With this we can improve testability and maintainability of our system.
- Unused methods should be moved to their desired class or should be identified as dead code and removed from the class to enhance maintainability.
- Class can be split into further smaller classes if it contains large number of methods. This will reduce the complexity and increase testability, stability and maintainability of the class.

In a nutshell we summarize our Factor/Criteria level findings in the below table. The up/down arrows indicates the parameter that should be increased or decreased in order to improve the overall quality of the system:

Table 17: Quality Summary (Factor/Criteria level)

| S.NO | Metric | MARF Class | GIPSY Class |
|---|---|---|---|
| 1 | Maintainability | ⬆ | ⬆ |
| 2 | Analyzability | ⬆ | ⬆ |
| 3 | Changeability | ⬆ | ⬆ |
| 4 | Stability | ⬆ | ⬆ |
| 5 | Testability | ⬆ | ⬆ |
| 6 | Complexity | ⬇ | ⬇ |
| 7 | Understandability | ⬆ | ⬆ |
| 8 | Flexibility | ⬆ | ⬆ |
| 9 | Efficiency | ⬆ | ⬆ |

**B.** At Class Metric Level

Here, we suggest some recommendations to improve the quality of the code at class metric level. From figure 37 we see that worst class from MARF leads/lags in 8 aspects when it comes to metrics while worst class from GIPSY leads/lags in 9 metrics level.

At the class level we not only check for metrics that are used to measure quality of the class but we also check for metrics which gives information pertaining to the interaction between multiple classes of the system.

For instance, in_bases gives the measure of the classes from which a class inherits. The higher the value the more will be the complexity. So ideally it should have a low value. This fundamental approach should be followed with cu_cdused, in_noc and cd_cdusers as well. The lower the value the more comprehendible the code is. Classes which involve these interactions can be measured using these metrics to determine the quality of the code.

Changes in one class should not impact other classes. Therefore, by reducing the interdependency we can minimize these variations in the above classes. Since both the classes are affected by these factors.

Furthermore, a source code should be well indented to enhance its readability and understandability. This metric stretches up to infinite level. By increasing the value of cl_comf and cl_comm we can indirectly increase the testability and modifiability of the code, as a user can easily analyze the code and implement.

Following table gives a brief metric related description for the two classes. The up/down arrows indicate that the corresponding metric should be increased/decreased to increase the productivity and quality of the application.

Table 18: Quality Summary (Metric level)

| S.NO | Metric | MARF Class | GIPSY Class |
|---|---|---|---|
| 1 | cl_comf |  | ⬆ |
| 2 | cl_comm |  |  |
| 3 | cl_data | ⬇ | ⬇ |
| 4 | cl_data_publ | ⬇ | ⬇ |
| 5 | cl_func | ⬇ | ⬇ |
| 6 | cl_func_publ | ⬇ | ⬇ |
| 7 | cl_line |  |  |
| 8 | cl_stat | ⬇ | ⬇ |
| 9 | cl_wmc | ⬇ | ⬇ |
| 10 | cu_cdused | ⬇ | ⬇ |
| 11 | cu_cdusers |  |  |
| 12 | in_bases | ⬇ | ⬇ |
| 13 | in_noc |  |  |

*C.B McCabe*

McCabe enables to deliver secure and more reliable software to the end users. McCabe uses advanced software metrics to identify measure and report on the complexity and quality of the code at the application and enterprise level. We have used it to indicate quality of methods and classes for both frameworks MARF and GIPSY.

**Case 1:** Quality of Methods [33][34][35]

Quality is calculated in terms of system's complexity.

**MARF**

**Average Cyclomatic Complexity, (v(G)):** McCabe proposes an original limit of 10 for the cyclomatic complexity. However, range is 0- infinity. It is also observed that adding a non-decision node does not change the cyclomatic complexity, but adding a decision node does. Different control structures impact differently on the complexity.

| Value Obtained | Interpretation |
|---|---|
| AVG V(G): 1.75 | Code has less number of decision nodes |

**Essential Complexity, (eV(G)):** Essential complexity is degree to which a code is unstructured. There is no particular limit to it, but poor module structure makes code hard to understand, thus hard to test. Therefore a lower value of eV(G) is favorable. Lower eV(G) means the code is well structured and organized. eV(G) value ranges from 1-infinity. Ideal threshold is 4.

The advantages of estimating Ev are:

- Quantifies the extent to which unstructuredness is present in a program.
- Helps in determining the quality of the code.

| Value Obtained | Interpretation |
|---|---|
| AVG ev(G): 1.20 | The code is well structured. |

**Module Design Complexity (iV(G)):** The module design complexity reflects the complexity of the module's calling patterns to its immediate subordinate modules. Thus a lower value of iV(G) is favorable for the code to be more testable and maintainable. And increase in the iV(G) will increase the design complexity, hence making it difficult to make the test cases. Values range from 0- infinity. Ideal threshold is 0.70.

| Value Obtained | Interpretation |
|---|---|
| AVG Iv(G): 1.57 | The code is complex to understand as it is not well designed. |

The table below summarizes the results discussed above.

**Table 19: MARF Complexity Summary**

| S.NO | Complexity Measure | Total value | Average Value |
|---|---|---|---|
| 1 | Average Cyclomatic Complexity | 3722 | 1.75 |
| 2 | Essential Complexity | 2556 | 1.20 |
| 3 | Module Design Complexity | 3332 | 1.57 |

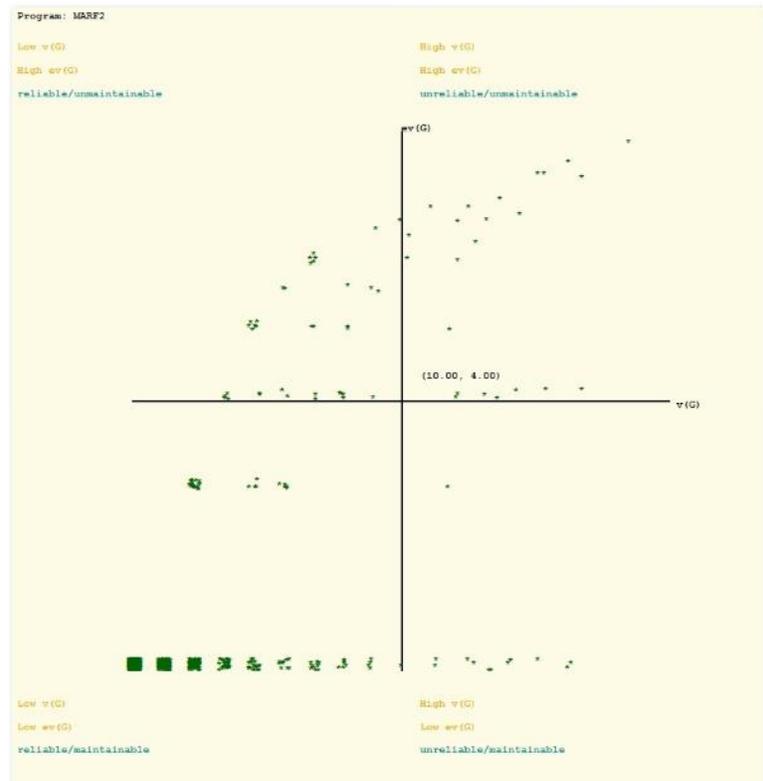

**Figure 35: Scatter Plot MARF**

The complexity levels are illustrated using the Scatter plot. Scatter plot is divided into 4 quadrants. The first two quadrants correspond to high complexity. The plot takes into account two values i.e. cyclomatic complexity and essential complexity.

Cyclomatic complexity is computed on the basis of control flow graph of the program. According to McCabe interpretation of complexity, a value between 1 to10 is ideal from simplicity and risk perspective while a value beyond 50 involved very high risk.

### GIPSY

**Average Cyclomatic Complexity, (v(G)):** If the value is high, there is higher risk due to difficulty in comprehension and testing. The commonly used threshold is 10. A high Cyclomatic complexity indicates decreased quality in the code resulting in higher defects that become costly to fix. Modules with V(G)>10 are at risk of reliability.

| Value Obtained | Interpretation |
|---|---|
| (v(G)) = 4.07 | It is a simple function without much risk. Probability of risk is low. |

**Essential Complexity, (ev(G)):** If the value is high, it leads to impenetrable code, which is higher risk due to difficulty in comprehension and testing. The commonly used threshold is 4. If the values exceeds it is difficult to maintain the code. Higher values indicate increased maintenance cost with decreased quality code. Module with ev(G)>4 are at risk of maintainability.

| Value Obtained | Interpretation |
|---|---|
| (ev(G))= 1.84 | The code is well structured. |

**Module Design Complexity(iv(G)):** The commonly used threshold is 0.70. However the minimum is 0 and maximum is 1.

| Value Obtained | Interpretation |
|---|---|
| (iv(G))= 3.01 | The code is not well designed. It is more complex. |

Summarized Tabular representation of data is described below:

**Table 20: GIPSY Complexity Summary**

| S.No | Complexity Measure | Total Value | Average Value |
|---|---|---|---|
| 1 | Average Cyclomatic Complexity | 24726 | 4.07 |
| 2 | Essential Complexity | 11172 | 1.84 |
| 3 | Module Design Complexity | 18729 | 3.01 |

The observed scatter plot for estimating complexity is represented by means of a graph.

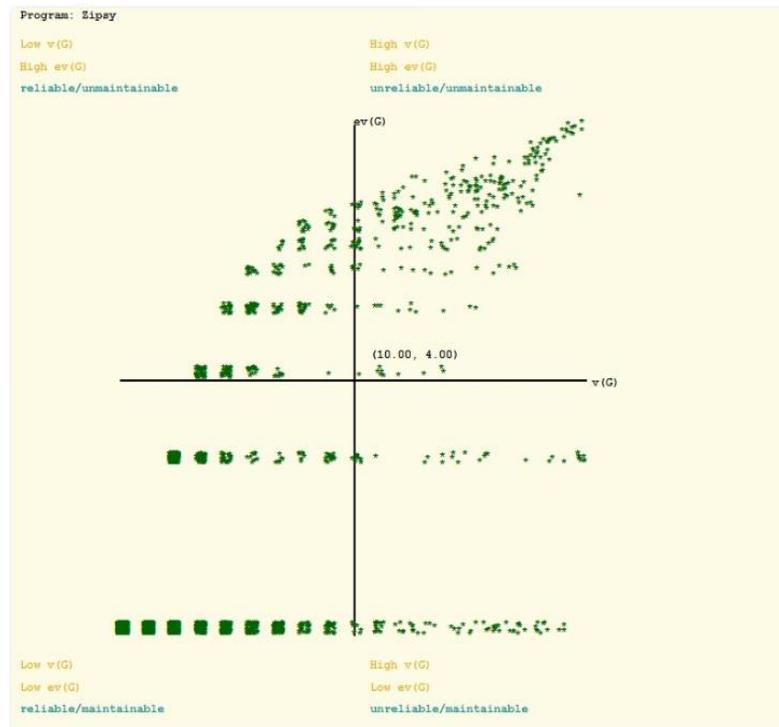

**Figure 36: Scatter Plot GIPSY**

**Case 2:** Quality of Classes [27]

High level metrics are used to calculate the average values. Each metric corresponds to a class characteristic.

**MARF**

**Average Coupling between Objects (CBO):** Class coupling is a measure of how many classes a single class uses. Good software design dictates that types and methods should have high cohesion and low coupling. High coupling indicates a design that is difficult to reuse and maintain because of its many interdependencies on other types. Range is from 0-infinity. An ideal value of 2 is optimal for CBO.

| Value Obtained | Interpretation |
|---|---|
| AVG: 0.17 | The code is less coupled hence more independent and easier to use in other parts of the application. |

**Weighted Methods per Class(WMC):** The WMC is a count of the methods implemented within a class or the sum of the complexities of the methods. The larger the number of methods in a class, the greater the potential impact on children; children inherit all of the methods defined in the parent class. Classes with large numbers of methods are likely to be more application specific, limiting the possibility of reuse. WMC values range from 0-infinity. Commonly used threshold is 14.

| Value Obtained | Interpretation |
|---|---|
| AVG: 11.41 | The code is easy to understand and modify as there are less methods. |

**Response for Class (RFC):** It gives the count of all methods that can be invoked in response to a message to an object of the class or by some method in the class. The larger the number of methods that can be invoked from a class through messages, the greater the complexity of the class. This in turn increases the testing and debugging of the class. Range is from 0-infinity. Ideal threshold is 100.

| Value Obtained | Interpretation |
|---|---|
| AVG: 16.43 | Lesser methods invoked in response to a message from some other method. Thus lesser complexity. |

**Depth of Inheritance Tree (DIT):** DIT measures the maximum level of the inheritance hierarchy of a class. A low number for depth implies less complexity but also the possibility of less code reuse through inheritance. A high number for depth implies more potential for code reuse through inheritance but also higher complexity with a higher probability of errors in the code. Range is from 1-infinity. Threshold is 7.

| Value Obtained | Interpretation |
|---|---|
| AVG: 2.14 | Code is lesser complex and less prone to fault. More code reusability |

**Number of Children (NOC):** NOC is the number of immediate subclasses subordinate to a class in the hierarchy. The greater the number of children, the greater is the reusability in the system. But, if a class has a large number of children, it may require more testing of the methods of that class, thus increase the testing time and effort. Range is 0-infinity. Threshold is 3.

| Value Obtained | Interpretation |
|---|---|
| AVG: 0.25 | Code reusability is less. Less test cases required. |

The table below summarizes the results obtained.

Table 21: MARF Metrics Summary

| S.NO | Metrics | Total value | Average Value |
|---|---|---|---|
| 1 | Average Coupling Between Objects | 31 | 0.17 |
| 2 | Weighted Methods per Class | 2066 | 11.41 |
| 3 | Response For Class | 2973 | 16.43 |
| 4 | Depth Of Inheritance Tree | 387 | 2.14 |
| 5 | Number Of Children | 45 | 0.25 |

**GIPSY**

We use these average values in order to determine the quality of classes.

**Average Coupling between Objects (CBO):** The commonly used threshold is 2 and minimum is 0. However increase in coupling limits the availability of class for reuse, and also results in greater testing and maintenance efforts.

| PM1 case study test results | Interpretation |
|---|---|
| CBO= 0.07 | The code is more independent and easier to use in another application. Maintenance of code is easy. |

**Weighted Methods per Class (WMC):** The commonly used threshold is 14 and minimum is 0. However increase in WMC lead to more faults. It limits the possibility of reuse. Increase in WMC increase the density of bugs and decreases the quality.

| PM1 case study test results | Interpretation |
|---|---|
| WMC= 10.54 | The code is easy to understand and modify. It is less fault-prone. It requires fewer efforts for testing and maintenance. |

**Response for Class (RFC):** The minimum value is 0 and can go up to N. The higher value indicates more faults. Classes with higher values are more complex and harder to understand. Testing and Debugging is complicated.

| PM1 case study test results | Interpretation |
|---|---|
| RFC= 12.62 | The code is quite easy to understand and testing and debugging of code is also easy. |

**Depth of Inheritance Tree (DIT):** The commonly used threshold is 5 and minimum value is 1. However increase in the value increase the faults. The code becomes more complex and harder to understand and modify.

| PM1 case study test results | Interpretation |
|---|---|
| DIT= 2.02 | The code is less prone to fault. It is easy to modify the code. It has the potential to reuse the inherited methods. |

**Number of Children (NOC):** The commonly used threshold is 3 and minimum is 0. However increase in the value increases the testing efforts. But high value of NOC indicates fewer faults this may be due to high reuse.

| PM1 case study test results | Interpretation |
|---|---|
| NOC= 0.21 | The code is likely to be more fault-prone. Reuse of code is less. However the code is stable and easy to modify |

Summarized results are described in the table below:

**Table 22: GIPSY Metrics Summary**

| S.No | Metrics | Total Value | Average Value |
|---|---|---|---|
| 1 | Average Coupling Between Objects | 38 | 0.07 |
| 2 | Weighted Methods per class | 6134 | 10.54 |
| 3 | Response For Class | 7347 | 12.62 |
| 4 | Depth Of Inheritance Tree | 1138 | 2.02 |
| 5 | Number Of Children | 120 | 0.21 |

*C.C Summary*

We have portioned the summary in two parts. Firstly, we discuss the measurement analysis for the classes discussed in the previous sections and Secondly, we discuss our opinion on the usage of both the tools namely McCabe and Logiscope which we used for report generation.

**A.** Measurement Analysis

Classes chosen by our team are:

- MARF.java and NeuralNetwork.java from the MARF Project.
- SimpleCharStream.java and GIPSYGMTOperator.java from Gipsy Project.

We define these classes under poorly developed category in comparison to the other classes in the application. Maintenance is the most expensive phase and also the most time consuming which demands developers to well understand the code and the initial level. Maintainability ensures easy and cost effective maintenance i.e. the software design should be easy to understand and modify. Maintainability can be broken down into further criteria's which are listed below:

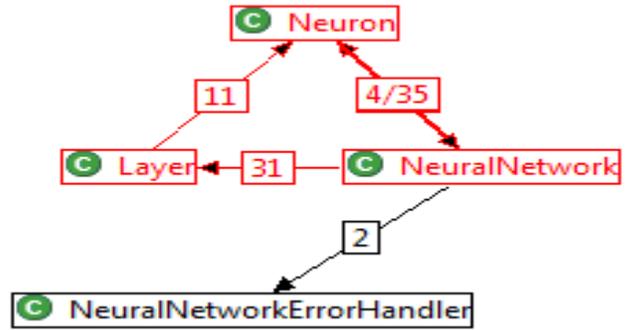

**Figure 37(a): Dependency Association**

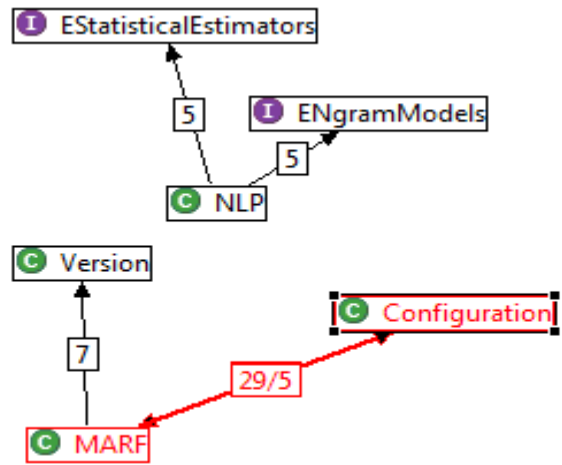

**Figure 37(b): Dependency Association**

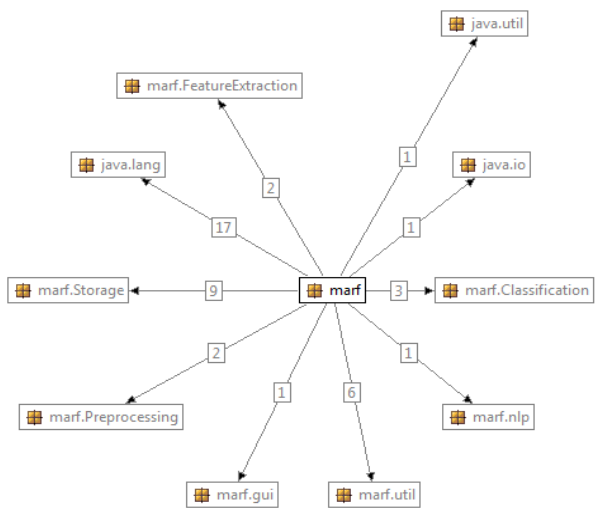

**Figure 38(a): Package Association**

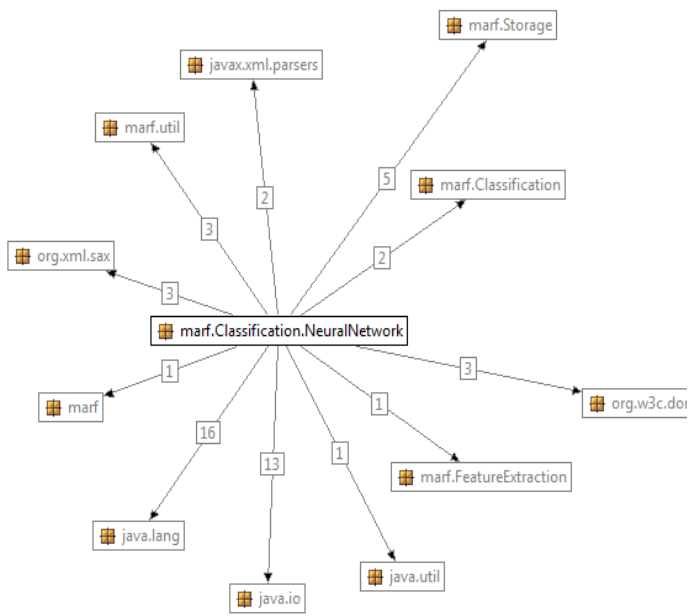

**Figure 38(b): Package Association**

The figures 37(a) and 37(b) shows the dependencies or association of Marf.java and NeuralNetwork.java with the other classes in their package.

NeuralNetwork.java has a very high coupling with Neuron.java, Layer.java and Marf.java has with Configuration.java and Version.java which implies any changes done to any of the coupled class can have an adverse effect on the associated class thereby making these classes vulnerable to errors which can impact the stability due to any intended or unintended changes.

Figures 38(a) and 38(b) shows the Package association of the native class i.e. Marf.java and NeuralNetwork.java with other packages in the application. There is high coupling between the native package and other packages which implies any change done to any class in the associated package will also have an effect on the native packages of other classes making the system unstable and making it difficult to test.

Some code level analysis is illustrated below to find certain poorly developed code constructs which leads to migration of errors. Eliminating these will help improving the Maintainability Factor.

**CODE SNIPPETS:**

NeuralNetwork.java

```
        catch(StorageException e)
        {
            e.printStackTrace(System.err);

            throw new ClassificationException
            (
                "StorageException while dumping/restoring neural
                e.getMessage(), e
            );

        catch(NumberFormatException nfe)
        {
            // TODO: throw an exception maybe?
            System.err.println("NumberFormatException: " + nfe.getMessage());
            nfe.printStackTrace(System.err);
        }
    }
    else
    {
        System.err.println("Unknown layer attribute: " + strAttName);
    }
```

The production code should not have printStackTrace statements or System.err statements as they lower the efficiency of the application and can eat up the memory space.

```
public void dumpXML()
throws StorageException
{
    dumpXML(getDefaultFilename());
}

/**
 * Overrides the default implementation of <code>restoreXML()</code>.
 * Merely calls <code>initialize()</code>.
 * @see marf.Storage.IStorageManager#restoreXML()
 * @see #initialize(String, boolean)
 * @since 0.3.0.6
 */
public void restoreXML()
throws StorageException
{
    initialize(getDefaultFilename(), false);
}

/**
 * @see marf.Storage.StorageManager#backSynchronizeObject()
 * @since 0.3.0.6
 */
```

In the above code we observe the overriding of a synchronized method which should be prohibited as it can cause error during maintenance phase and could be difficult to understand.

```
DocumentBuilder oBuilder = oDBF.newDocumentBuilder();
OutputStreamWriter oErrorWriter = new OutputStreamWriter(System.err, OUTPUT_ENCODING);

oBuilder.setErrorHandler(new NeuralNetworkErrorHandler(new PrintWriter(oErrorWriter, true)));

Debug.debug("Parsing XML file...");
Document oDocument = oBuilder.parse(new File(pstrFilename));
```

The above code section shows the creation of an object of PrintWriter but is not closed explicitly by the developer.

For MARF:

```
/**
 * Must never be instantiated or inherited from...
 * Or should it be allowed?
 */
private MARF()
{
}
```

In this we found a blank definition of the constructor. It can create confusion or can create errors if a call is given from the associated classes to its constructor. And declaring it as private can further complicate it because it will prevent from instantiating the class. Hence it's better to remove this code construct.

For GIPSYGMTOperator:

```
private void loadView()
{
}
/**
 * Continue the simulation if we've loaded a saved one.
 */
public void continueSimulation()
{
}

/**
 * Pops up a File Dialog and Load a saved simulation based
 * selected in the Dialog Box.
 *
```

```
/**
 * Save a running simulation to a binary file.
 */
public void saveSimulation()
{
}

/**
 * Handles the buttons's events.
 *
```

The above code snippets show Empty Method declaration of the method in the class which is not being used. Hence this should be removed from the class.

```
private void startInstance()
{
    try
    {
        // -- Start the GMT first.
        System.out.println("Starting the GMT...");
        oGIPSYEntityController.startGMTNode();
        System.out.println("GMT started successfully...");
        oGIPSYEntityController.allocateTier();
    }
    catch (Exception e)
    {
        System.err
            .println("An error occured while trying to start instances.\n Error"
                + e.getMessage());
    }
}
```

```
private boolean isMapLoaded()
{
    boolean isValid = false;

    GlobalInstance oSingleInstance = GlobalInstance.getInstance();
    if (oSingleInstance.getGIPSYInstancesList().size() > 0
            && oSingleInstance.getGIPSYNodesList().size() > 0
            && oSingleInstance.getGIPSYTiersList().size() > 0)
    {
        isValid = true;
    }

    return isValid;
}
```

The methods shown in above are declared and implemented in the class but are nowhere used within the class. These should be removed from this class and moved to a more appropriate class. These are increasing the complexity of the class and making it difficult to understand.

```java
private void enablePanels()
{
    try
    {
        createPanels();
        loadListPanels();
        enableButtons();
        bIsRunning = true;
    }
    catch (Exception e)
    {
        System.err.println(e.getStackTrace());
    }
}
```

The above code shows the use of System.err or System.Out statements in the production code which should be avoided as it can hamper the efficiency and can be a challenge in maintainability later on.

For SimpleCharStream :

- It contains large number of methods hence increasing the complexity of the class and making it difficult to test, maintain and analyse the code.It should be further broken into smaller classes.
- The default constructor for the class is not present. It should be declared in the class to make it more maintainable for future purpose.

**B.** Tool Experiences

Logiscope and McCabe tools provide a means for measuring the quality of a system. McCabe helps to deliver better and secure software. It uses advanced software metrics to identify measure and report the complexity and quality of the code at the application and enterprise level while logiscope measures the quality of the code in terms of software metrics, which indicates its complexity, testability, understandability and its description. Both the tools use different means to provide the quality related information. [35]

Although, these tools are used for more or less similar purposes but still there exists certain differences.

Logiscope is not multi-lingual as it does not provides multi-language support which limits its usage to only few language specific projects. On the other hand McCabe IQ analyzes many programming languages on any platform which enables the organization to use the same tool for all the company projects.

Logiscope tool allows us to find the only latent bugs, complex errors, code similarities and poorly structured code that make our software complex to test and maintain. There is no advanced visualization in this tool. Therefore, it only finds the bugs and errors whereas McCabe uses the concept of advanced visualization which presents detailed color-coded graphical display through which we can visualize our code and unravel the logic, architecture and designs.

Fixing bugs, doing maintenance or testing software is indispensable for engineers and managers. It enables us to visualize metrics and parameters in the context of the entire application.

Logiscope provides a good technology but not the best. It tackles only the software quality head-on providing a comprehensive suite of highly customizable static and dynamic testing tools. With logiscope we need supportive tools to perform specific tasks but McCabe does not have any such disadvantage.

It mainly measures the cyclomatic complexity of the code. Cyclomatic complexity of a source code directly measures the number of linearly independent paths through the programmer source code. It uses the control flow graph of the program. It also applied to individual functions, classes and modules within a program.it uses basis path testing strategy to test the each linear independent path.

In a nutshell, McCabe outshines logiscope when it comes to performance and ease of use but still we cannot neglect the importance of logiscope to tackle software quality head-on.

*D. Design and Implementation with JDeodorant and MARFCAT*

*D.A Overview*

In this section we describe code smell detectors namely JDeodorant and MARFCAT, which enable programmers and developers to determine opportunities for improvement. Smell detector basically refers to a tool that identifies code smells automatically. Code smells are the flaws in system's design or overall structure. We have used these tools to detect smells in the source code of MARF and GIPSY. Refactoring technique is used for the improvement purpose. Refactoring is essential to make a system more understandable and maintainable. A brief description regarding usage and features of these detectors is described below.

*D.A.A JDeodorant*

JDeodorant is an Eclipse plug-in which is used to identify design problems in different software's. It is capable of detecting and resolving bad smells. JDeodorant makes use of a variety of novel methods and techniques in order to identify code smells and suggests the appropriate refactoring to resolve them. It can also be referred as a refactoring tool.[29]

Following classification describes the types of smells detected by JDeodrant:

- **God Class -** A god class refers to an object, which is used to control multiple objects in a system such that it becomes a class that does everything.
- **Long Method** - This smell refers to a function or procedure that has become too large.
- **Type Checking** - It occurs either when attribute of a class represents state or there is a conditional statement.
- **Feature Envy** - This smells refers to a class that excessively accesses the methods of other classes.

Each of these smells can be resolved by various refactoring strategies:

- Move Method refactoring is used to overcome Feature Envy problems.

- Type Checking problems are resolved by appropriate Replace Condition with Polymorphism and Replace Type code with State/Strategy refactoring.

- Long Method problems are resolved by appropriate Extract Method refactoring.

- God Class problems are resolved by appropriate Extract Class refactoring.[28]

The primary advantage of using JDeodrant is that it completely transforms expert's knowledge to an automated process. All other advantages of using JDeodrant are summarized below:

- Pre-Evaluation of the effect for each suggested solution.
- User guidance in comprehending the design problems.
- User friendliness (one-click approach in improving design quality).
- Java Language Specification 4 (Java 7)

Even though JDeodrant offers multiple features but it has two major flaws:

**Language Dependency** - It supports only Java language.

**Error Detection Capacity** - The detection power is only limited to four smells.

*D.A.B MARFCAT*

MARFCAT is MARF (Modular Audio Recognition Framework) based code analysis tool. It can be used for following functions:

- Machine learning techniques.
- Categorizing
- NLP techniques
- Finding vulnerabilities in the code.
- Finding deficiencies related to security and software engineering.

MARFCAT does not have any concern with the language, byte code, source code or the binary being examined.

It is used for machine learning, NLP and spectral techniques. We can divide machine learning into two categories:

- **CVE (Common Vulnerabilities and Exposures)**

  Vulnerability is defined as a flaw in a software system through which a hacker can access the system or network while Exposures are the configuration issues and flaws in a software through which hacker can access the useful information and hack the system or network.

- **CWE (Common Weakness Enumeration)**

  CWE provides software security tools and services to find the weakness in the code and solve the problem related to design and architecture.

  One of the disadvantages is that the identification of useful CVE and CWE in non CVE & CWE case needs a firm knowledge base.[35]

**Methodology of MARFCAT**

The steps referred below describe the MARFCAT methodology.

- Firstly, compile meta-XML files from the CVE reports (line numbers, CVE, CWE, fragment size etc.). This becomes an index mapping of CVEs to files and locations within files
- On the basis of the meta files educate the system to construct the knowledge base.
- Examine the training data for the same case and identify the best algorithm combinations for the task.
- Analysis on the testing data for the same case without the annotations as a sanity check.

- Examine the testing data for the fixed case of the same software.
- Assessment on the testing data for the general non-CVE case.

The advantages of using MARCAT are:

- **Language Independency** - MARFCAT is not limited to Java, rather it offers capabilities regardless of the underlying language.
- **Accuracy and Speed** – It is fast and efficient in terms of usage.
- **Pre-Scanning** – It can be used to quickly pre-scan the projects for analysis purpose.
- **Vast and Descriptive** – Large amount of algorithms and their combinations are available.
- **Easy Dealing** - It can easily deal with the changed code and the code which is used in other project.
- **Testing** - It can autonomously learn the large knowledge base to test known and unknown cases.

Despite the fact MARFCAT offers plenty of features but its limitations cannot be negated.

- Accuracy achieved using MARFCAT depends on the quality of knowledge base.
- Separating wavelet filter performance badly affects the precision.
- No parsing, slicing, semantic annotations, context, locality of reference.
- Graphical User interface is not quite good.
- In order to detect the CVE/CWE signatures in non-CVE cases large knowledge bases are required.

*D.B Design and Implementation*

This section is divided into two categories. Firstly, we discuss the metrics design and implementation using JDeodorant and Secondly, we use MARFCAT for a vulnerable code scan.

**PART A:** Using JDeodoramt

We have used this well-known code detector in our analysis. On the basis of our analysis in section B.I, we concluded that design properties greatly influence the quality of the code. Encapsulation is one the properties of Object-Oriented systems, which is directly associated with system's flexibility, understandability and effectiveness. Similarly, inheritance is linked with system's extendibility and effectiveness. Therefore, we prioritize two metrics from the MOOD and QMOOD set which directly has an impact on these design properties.

Thus, we finalize our metrics as:

- AHF
- DIT

**AHF** - AHF stands for Attribute Hiding Factor. This metric is used to measure how metrics are encapsulated in the classes of a system. Ideally, AHF should have a high value as it is expected that all the attributes of a class should be hidden.

It is defined as the ratio of sum of all attributes to the total number of attributes and is used to measure encapsulation.

$$AHF = \frac{\sum_{i=1}^{TC}\sum_{m=1}^{Ad(ci)}(1 - V(Ami))}{\sum_{i=1}^{TC} Ad(Ci)} \qquad 12$$

where,

$$V(Ami) = \frac{\sum_{j=1}^{TC} \text{visible}(Ami, Cj)}{TC - 1}$$

and ,

visible(Ami,Cj) = 1 if $j \ne$ Cj may call Ami

0 otherwise

**DIT** - DIT measures the depth of inheritance tree i.e. the maximum level of inheritance hierarchy of a class. It is based on directly counting the level of a class from the root of inheritance tree.

The higher the value of DIT, the more difficult it is to understand and maintain a class. Also, classes become more susceptible to errors due to design violations.

Apart, from the above metrics, we have also covered NOC, WMC and NOM. The aforesaid metrics are categorized under Inheritance level, Class level and Method level respectively.

- **Inheritance Level**

    **NOC** - NOC denotes the number of children. It represents the number of immediate subclasses of a class. It is obtained by counting the sub classes of a class. The higher the value the higher reusability it offers but it could also lead to difficulty in understanding and modifying the code. It also increases the effort required in testing.

- **Class level**

    **WMC** - Weighted method per class gives the measure of system's complexity. It is the sum of individual complexities of all the methods. Thus it indirectly relates to the effort required to develop and maintain. The larger the number of methods, it is more likely to be complex as compared to a class with less number of methods.

    $$WMC = \sum_{i=1}^{n} Ci \qquad 13$$

    where, Ci refers to the individual complexity.

    **LCOM** – LCOM gives lack of cohesion of methods. It considers methods as cohesive of they share atleast one common instance variable. For example if we consider P as the set of methods pairs which does not share any instance variable in common and Q is defined as the set of method pairs which share the attributes in common. Thus, we can define LCOM in terms of P and Q.

    LCOM =  |P| - |Q|   if |P| > |Q|        14
            0   otherwise

    The above is the CK definition of LCOM. A high LCOM values implies lack of cohesion and low value indicates high cohesion. Ideally, a class should have high cohesion and low coupling.

- Method level

    **NOM** – Number of Methods gives the count of number of methods in a class. It directly impacts the complexity. Greater number of methods increases complexity.

We have implemented all these metrics under the metrics package of JDeodorant which can be viewed in the file attached.

In order to validate the code we have generated some dummy test cases to calculate the values of above metrics. Table below gives the values obtained using metrics and it is compared with the actual values.

tests.suite1(DIT, NOC)

| Metric/ Class Name | NOC | NOM | AHF | WMC | DIT | LOC |
|---|---|---|---|---|---|---|
| A | 2 | X | X | X | 0 | X |
| B | 0 | X | X | X | 1 | X |
| C | 0 | X | X | X | 2 | X |
| D | 1 | X | X | X | 1 | X |
| SuperC | 0 | X | X | X | 0 | X |

tests.suite2(NOM)

| Metric/ Class Name | NOC | NOM | AHF | WMC | DIT | LOC |
|---|---|---|---|---|---|---|
| A | X | 2 | X | X | X | X |
| B | X | 3 | X | X | X | X |
| SuperC | X | 1 | X | X | X | X |

tests.suite3(WMC)

| Metric/ Class Name | NOC | NOM | AHF | WMC | DIT | LOC |
|---|---|---|---|---|---|---|
| A | X | X | X | 2 | X | X |
| B | X | X | X | 4 | X | X |
| C | X | X | X | 3 | X | X |
| SuperC | X | X | X | 1 | X | X |

tests.suite4(AHF)

| Metric/ Class Name | NOC | NOM | AHF | WMC | DIT | LOC |
|---|---|---|---|---|---|---|
| A | X | X | 173.4375% | X | X | X |
| B | X | X | 296.875% | X | X | X |
| SuperC | X | X | 321.875% | X | X | X |

Next, we run our metric implementations on metric implementations themselves to determine the quality of our own implementation.

Table below summarizes the metric implementations on metric implementations itself.

**Table 23: Metric Implementation**

| S.No | Metric | File Name | Metric Value |
|---|---|---|---|
| 1 | LCOM | LCOM.java | 3 |
| 2 | AHF | AHFMetric.java | 24.46 |
| 3 | DIT | DITMetric.java | 0 |
| 4 | NOC | NOCMetric.java | 0 |
| 5 | WMC | WMCClass.java | 5 |
| 6 | NOM | NOMClass.java | 2 |

Finally, we run these metrics on the two worst quality classes we determined in Measurement Analysis section of C.C.

**Table 24: Metric Implementation on Worst Code**

| Files | LCOM | AHF | DIT |
|---|---|---|---|
| Marf.java | 1260 | -76.08% | 0 |
| NeuralNetwork.java | 175 | -50.02% | 2 |
| SimpleCharStream.java | - | -16.14% | 0 |
| GIPSYGMTOperator.java | - | -16.34% | 1 |

| Files | NOC | WMC | NOM |
|---|---|---|---|
| Marf.java | 0 | 21 | 57 |
| NeuralNetwork.java | 0 | 61 | 26 |
| SimpleCharStream.java | 0 | 25 | 29 |
| GIPSYGMTOperator.java | 0 | 27 | 36 |

From our analysis it is evident that these four classes are worse in terms of their overall quality. For computing the size of problematic classes we use the LOC as a measure.

We have used cloc to calculate the lines of code in each package to which the files belong.

Below outputs are generated for MARF and GIPSY files.

**Table 25: LOC Measure**

| Package Name | File Name | Package Size | File Size |
|---|---|---|---|
| MARF | MARF.java | 24929 | 750 |
| NeuralNetwork | NeuralNetwork.java | 1657 | 375 |
| util | SimpleCharStream.java | 622 | 355 |
| ui | GIPSYGMTOperator.java | 4756 | 644 |

Here, package size represents the LOC of entire package whereas; file size represents LOC of the problematic file.

In order to evaluate the percentage in terms of size of the problematic vs less problematic classes with the package, we refer to the data listed in above table.

**Case 1**: MARF

```
              Package MARF
-----------------------------------------------------------
Language      files     blank    comment      code
-----------------------------------------------------------
Java           200      6692      21329      23954
make            38       417        230        975
-----------------------------------------------------------
SUM:           238      7109      21559      24929
-----------------------------------------------------------

                Marf.java
-----------------------------------------------------------
Language      files     blank    comment      code
-----------------------------------------------------------
Java             1       254        930        750
-----------------------------------------------------------
```

**Table 26: Package MARF Size %**

| Package MARF | |
|---|---|
| Total LOC ( including subdirectories ) | 24929 |
| Total LOC ( excluding subdirectories ) | 1153 |
| MARF.java LOC | 750 |
| Size % Problematic ( including subdirectories ) | 3% |
| Size % Problematic ( excluding subdirectories ) | 65.04% |
| Size % Less Problematic ( including subdirectories ) | 97% |
| Size % Less Problematic (excluding subdirectories ) | 34.96% |

In this case the problematic size percentage is negligible because of the directory structure of MARF. MARF contains a total of 200 Java files. Thus, marf.java constitutes to only 3% of the total but excluding the directory structure there are only 3 Java files and therefore, the percentage becomes 65%.

**Case 2:** NeuralNetwork

```
           Package NeuralNetwork
-----------------------------------------------------------
Language      files     blank    comment      code
-----------------------------------------------------------
XML              4        17          5       1044
Java             3       228        101        561
Perl             1        29          5         52
-----------------------------------------------------------
```

```
SUM:             8        274       111      1657
--------------------------------------------------
                   NeuralNetwork.java
--------------------------------------------------
Language        files     blank    comment    code
--------------------------------------------------
Java              1        128       34       375
--------------------------------------------------
```

**Table 27: Package NeuralNetwork Size %**

| Package NeuralNetwork | |
|---|---|
| Total LOC | 1657 |
| NeuralNetwork.java LOC | 375 |
| Size % Problematic | 22.63% |
| Size % Less Problematic | 77.37% |

In this case the problematic size percentage is nearly 24%. Considering only very few files in the package, NeuralNetwork.java increases the complexity levels.

**Case 3:** Util

```
                    Package Util
--------------------------------------------------
Language        files     blank    comment    code
--------------------------------------------------
Java             39        945      2417      4756
--------------------------------------------------
SUM:             39        945      2417      4756
--------------------------------------------------

                  SimpleCharStream.java
--------------------------------------------------
Language        files     blank    comment    code
--------------------------------------------------
Java             39        945      2417      4756
--------------------------------------------------
SUM:             39        945      2417      4756
--------------------------------------------------
```

**Table 28: Package Util Size %**

| Package Util | |
|---|---|
| Total LOC | 622 |
| SimpleCharStream.java LOC | 355 |
| Size % Problematic | 57.07% |
| Size % Less Problematic | 42.93% |

In this case the problematic size percentage is greater than 50%. It is highly complex when it comes to other classes in the package.

**Case 4:** Ui

```
                    Package Ui
--------------------------------------------------
Language        files     blank    comment    code
--------------------------------------------------
Java             39        945      2417      4756
--------------------------------------------------
SUM:             39        945      2417      4756
--------------------------------------------------

                 GIPSYGMTOperator.java
--------------------------------------------------
Language        files     blank    comment    code
--------------------------------------------------
Java              1        104       171      644
--------------------------------------------------
```

**Table 29: Package Ui Size %**

| Package Ui | |
|---|---|
| Total LOC | 4756 |
| GIPSYGMTOperator.java LOC | 644 |
| Size % Problematic | 13.54% |
| Size % Less Problematic | 86.46% |

Similarly, the problematic size percentage is nearly 14%. On comparing this class with other classes of the package, we infer that the problem percentage is quite low.

Next we compare these worst quality classes with other classes within the same package.

MARF.java has a WMC value 21 while NeuralNetwork.java has a WMC value of 61. Although, individually these two classes are highly complex and present a risk to system's reliability but when the overall complexity of the package is compared with the complexity of these classes, it does not create much effect.

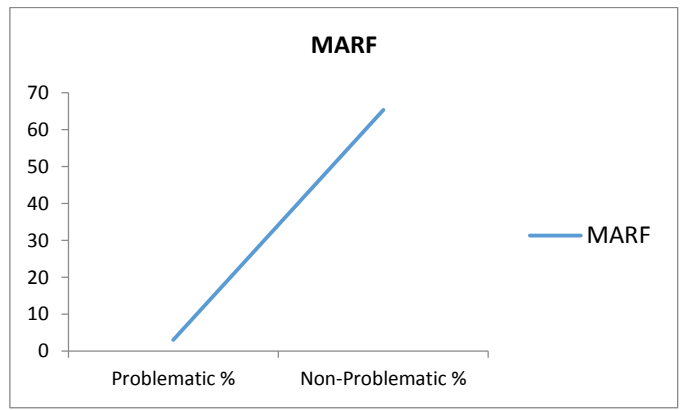

**Figure 39: Problematic vs Non-Problematic MARF**

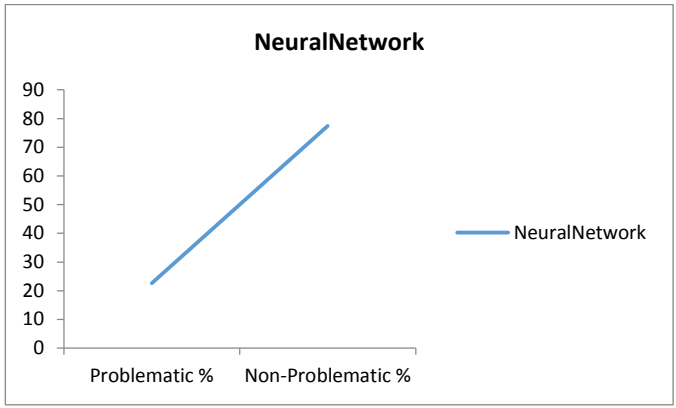

Figure 40: Problematic vs Non-Problematic Neural Network

Similarly, when problematic classes of GIPSY are compared with the less problematic classes of the same package, we observe that WMC of GIPSYGMTOperator.java is 27. Using Table 24 we can examine the metric values for these classes. The commonly used threshold for WMC is 14 and minimum is 0. However increase in WMC leads to more faults. It limits the possibility of reuse. Increase in WMC increases the density of bugs and decreases the quality as already discussed in the previous section.

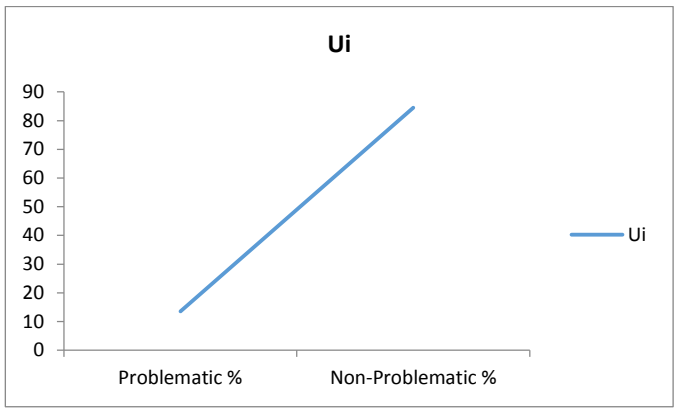

Figure 41: Problematic vs Non-Problematic Ui

All the classes do not create much impact on the performance of the package as a whole with an exception to class SimpleCharStream.java in the Util package. Here, the problem percentage is dominating within the package. This indicates that the poorest class is the dominant class in the package. The curve below shows a decreasing trend.

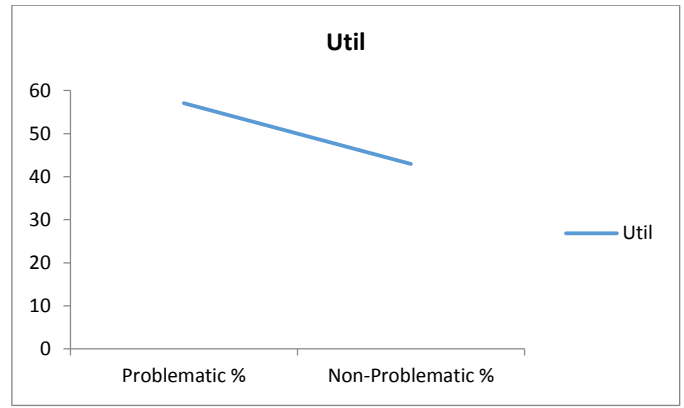

Figure 42: Problematic vs Non-Problematic Util

### PART B: Using MARFCAT

In this regard, attempt was made to do a code scan for vulnerabilities. The main advantage of using this approach is that it can be applied to any target language. Following steps were followed as part of the scan:

- All the scripts containing the executables and Jar files for both MARF and GIPSY were placed in a directory.
- The data from corresponding files were extracted and placed in the right folder.
- User was granted the appropriate privilege to execute the scripts by changing the permissions and,
- Finally, the test-quick-gipsy-cve and test-quick-marf-cve scripts were executed to do a code scan for GIPSY and MARF respectively.

After the scan log files are generated for all the test cases. The output of a log file looks something like this:

```
                File: test-
cases/marf/marf/src/marf/Classification/Distance/HammingDist
.java
              Path ID: 12
               Config: -nopreprep -raw -fft -cheb -flucid
      Processing time: 0d:0h:0m:0s:8ms:8ms
          Subject's ID: 44
    Subject identified: CVE-2008-5515
ResultSet: [suppressed; enable debug mode to show]
FileItem: strFileID: [12]
strPath: [test-
cases/marf/marf/src/marf/Classification/Distance/HammingDist
.java]
strFileType: [Perl5 module source text]
bEmpty: [false]
oLocations: [[[]]]

        Second Best ID: 39
      Second Best Name: CVE-2009-0783
             Date/time: Wed Jun 25 23:17:11 EDT 2014
Outcome o (classifier-specific): 917.9274110493719
         Distance threshold: 0.1
         Computed raw P = 1/o: 0.0
         Warning to be reported: false
         Computed normalized P: 0.0
```

**Figure 43: MARFCAT Logs**

Using the log we found that none of the test cases correspond to the Warning to be reported as true.

Therefore, we infer that neither the GIPSY nor the MARF system is affected by the vulnerabilities. For each of the class the outcome value was much above the threshold level. The default threshold is set to 0.1 which indicates any deviation below or equal to this level is problematic. The general rule for the basis of this decision is:

$$o <= 0.1$$

The code in the classes is vulnerable if this outcome value O value is less than 0.1.In such a situation "Warning to be reported" parameter becomes true.

These results are in contrast with the Tomcat 5.5.13 where the code was found to be vulnerable when the same analysis was performed. These vulnerabilities are expressed as good or bad in terms of code structure, its documentation and comments.

But in the current scenario of MARF and GIPSY these issues are not observed. The average of outcomes is:

MARF o(avg) = 984.6145193
GIPSY o(avg) = 1068.924959

The average of the outcome sample lies nowhere close to the threshold level and the distance gap between the individual classes' outcome and the threshold is very high.

Thus we conclude that MARF and GIPSY cannot be expressed in a way Tomcat was expressed through this vulnerability scan and none of the files were found with outcomes greater than the standard threshold level.

*D.C Summary*

To sum up, both JDeodorant and MARFCAT are powerful tools for identifying bad smells from code and resolve them by performing refactoring. JDeodorant is a much targeted tool that focuses on finding four specific bad smells namely Large Class (called God Class), Long Method, Duplicated Code and Feature Envy. It provides high user friendliness by giving refactoring advice to the users.

On the other hand, MARFCAT is language independent, accurate and fast. It can easily deal with large amount of code, which is used in other system. Interpretation of Graphical representation for code analysis is somewhat hard for humans. It is not as precise and accurate as compared to JDeodorant. Since, there is no parsing we cannot trace the path, slicing, semantic annotations, context, and locality of reference. To augment the scalability we convert the MARFCAT stand-alone application to a distributed one.

In the current context we used JDeodorant for metric implementations while MARFCAT for smell detection and removal.

*D.C.A Results*

Same analysis was made using McCabe to verify the results obtained using metric implementations. Below table gives the details of values obtained using McCabe.

**Table 30: McCabe results**

| MARF.java | | |
|---|---|---|
| **Metric** | **McCabe** | **Metric Implementation** |
| NOC | 0 | 0 |
| DIT | 1 | 0 |
| WMC | 58 | 21 |
| NOM | - | 57 |
| AHF | - | -76.08% |
| LCOM | 97 | 1260 |

| NeuralNetwork.java | | |
|---|---|---|
| Metric | McCabe | Metric Implementation |
| NOC | 0 | 0 |
| DIT | 3 | 2 |
| WMC | 27 | 61 |
| NOM | - | 26 |
| AHF | - | -50.02% |
| LCOM | 93 | 175 |

| SimpleCharStream.java | | |
|---|---|---|
| Metric | McCabe | Metric Implementation |
| NOC | 0 | 0 |
| DIT | 1 | 0 |
| WMC | 38 | 25 |
| NOM | - | 29 |
| AHF | - | -16.14% |

| GIPSYGMTOperator.java | | |
|---|---|---|
| Metric | McCabe | Metric Implementation |
| NOC | 0 | 0 |
| DIT | 2 | 1 |
| WMC | 37 | 27 |
| NOM | - | 36 |
| AHF | - | -16.34% |

From the above data it is evident that the results obtained using metric implementations are nearly similar. Minimum DIT value in McCabe starts from 1 while in the metric implementation the root of inheritance tree starts from 0. This is the reason of the value margin between the results obtained using McCabe and metric implementations.

In the above implementation we observe the negative value of AHF which indicates that the value of visible (Ami,Cj) is very large which in turn results in the negative value in the final computation. However, ideal value of AHF (100%) is desired. Ideal value of AHF ensures that all the attributes of the class are private.

However, we see a significant difference in WMC and LCOM values obtained using McCabe and metric implementations. This is a result of different complexity measurement criteria used in the two methods. But both the methods indicate a higher value because complexity which pose a risk to the system.

*D.C.B Analysis and Interpretation*

**MARF.JAVA**

The NOC count suggests lack of inheritance, thus less code reusability. The DIT count suggests that there is low complexity, no inheritance used in the class. Hence there is no code reusability in that class. The low RFC count suggests lesser complexity and lesser effort for testing and debugging. The high count for WMC suggests that the class has more methods hence more complex and thus less reusable. The null value of CBO suggests that there is no coupling between the classes. The LCOM value suggests that there is more cohesion between the methods. And less coupling and high cohesion is the sign of a good design. The scatter plot obtained using McCabe gives the complexity indication of this class.

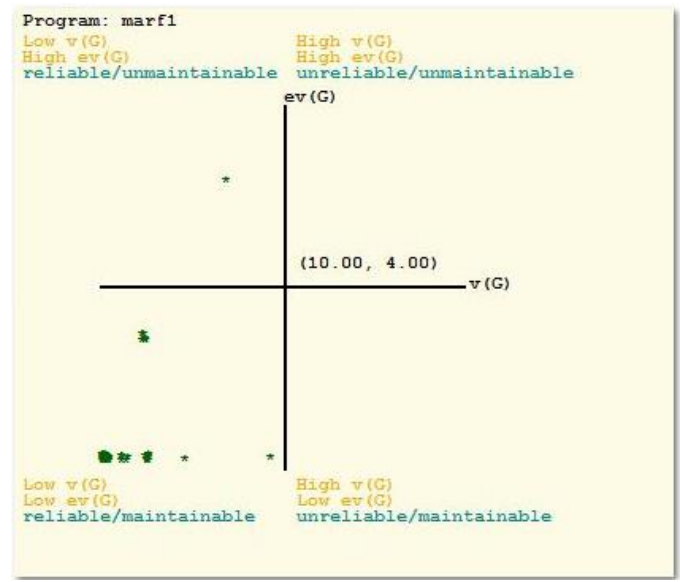

**Figure 44: MARF.java Scatter Plot**

**NeuralNetwork.java**

The NOC suggests that there is no reusability due to absence of inheritance. Hence there will be fewer test cases. The DIT value suggests low complexity and less reusability of code. The low RFC suggests lesser complexity. Thus there will be lesser test cases. The value for WMC is high, thus suggesting that the methods are more complex and require more testing and debugging. The CBO value shows absence of coupling. The LCOM value is again high and thus ideal for a good design as the cohesion in high. The scatter plot obtained using McCabe gives the complexity indication of this class.

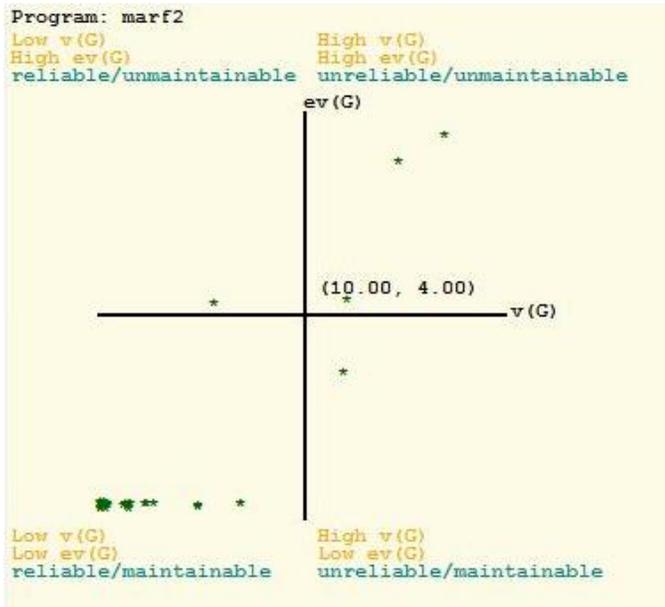

**Figure 45: NeuralNetwork.java Scatter Plot**

**SimpleCharStream.java**

The NOC metric suggests that there is no child extending from the class. There no reusability of code hence less complex and lesser testing is required. The DIT value also suggests that there is low complexity and lesser reusability of code. The RFC suggests that there is lesser complexity and hence less testing and debugging is required. The WMC is high that suggests that there is more complexity in the methods hence hence less reusability. The CBO count suggests that there is low coupling and high LCOM suggests high cohesion, which is again effective for a good design. The scatter plot obtained using McCabe gives the complexity indication of this class.

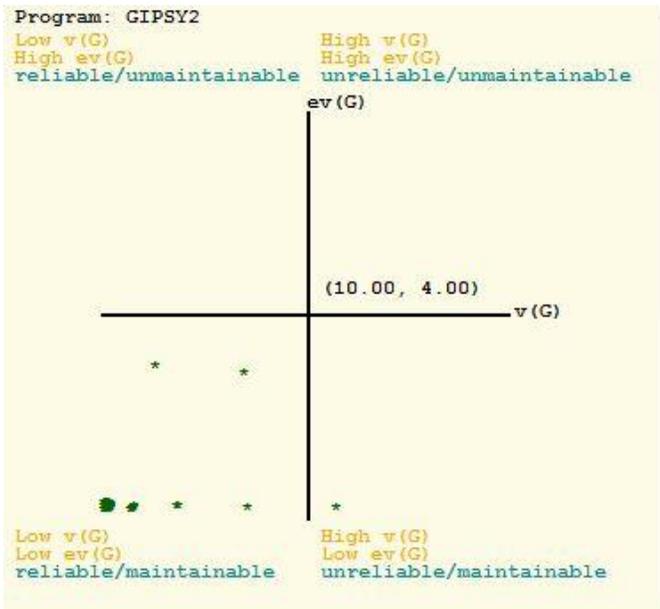

**Figure 46: SimpleCharStream.java Scatter Plot**

**GIPSYGMTOperator.java**

The NOC is suggests that there is no complexity, no reusability and no inheritance implemented in the class. The DIT value suggests that there is less complexity. The RFC count is very low, and suggests that there is less complexity hence it takes lesser effort for testing and debugging. The high WMC value suggests that the methods are more complex and thus less reusable. The CBO value suggests that there is no cohesion between classes. The high value of LCOM suggests that there is high cohesion. Thus the class is good as there is less coupling and more cohesion. The scatter plot obtained using McCabe gives the complexity indication of this class.

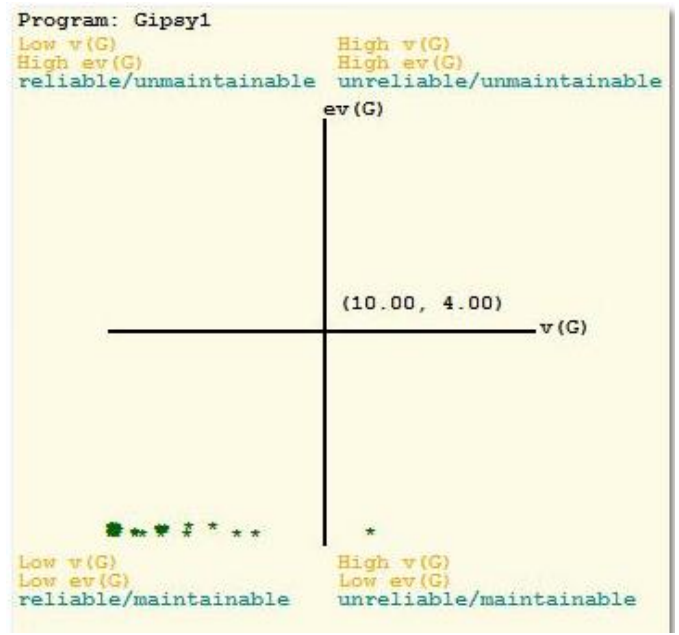

**Figure 47: GIPSYGMTOperator.java Scatter Plot**

*E. Conclusion*

In our entire study we analyzed various aspects of MARF and GIPSY with special concern to evaluate the quality of the code. Initially we interpreted that attributes like extensibility, modifiability, maintainability, integrity and adaptability are supported by MARF. Similarly, GIPSY was found to be scalable and flexible in this regard. Although GIPSY offers more flexibility than MARF but this feature increases its complexity. The source code of GIPSY is far more difficult to understand and comprehend as compared to MARF. The complexity relationship between MARF and GIPSY can be illustrated by below curve:

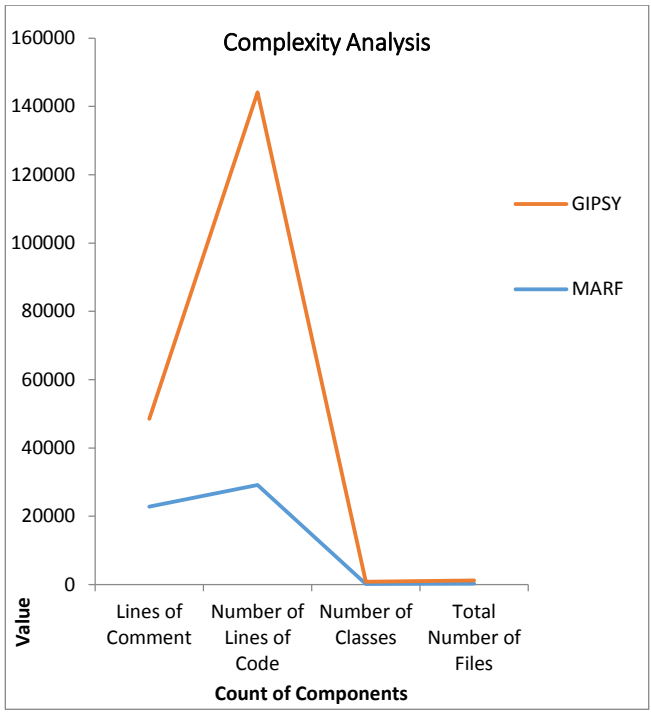

**Figure 48: Complexity Analysis MARF & GIPSY**

The above curve provides the complexity relationship considering Lines of Code as a measure

However, we observe from Figure 21 that maintainability levels of both the applications in consideration of this study are closely related to each other.

Moreover, a relative study of worst quality code of MARF and GIPSY manually and using the metrics described in the above sections reveals that worst quality code of GIPSY is even worse than the worst quality code of MARF.

Therefore, we identify a need to refactor or use some means to improve the quality of GIPSY and reduce its complexity.

## V. Contribution Summary

| Serial No | Name | Research Paper Summary | Task Summary |
|---|---|---|---|
| 1 | Suman Alungh | [15], [16], [17] | Linux Analysis, Sloccount, Report Consolidation, OSS case study, Summary Analysis, McCabe IQ, Logiscope, Review, Comparison Analysis, Metric Implementation |
| 2 | Chanpreet Singh | [18], [19], [20] | Linux analysis, Sloccount, OSS case study, Summary Analysis, Review, Metric level Analysis, Comparison Analysis, Report Consolidation, Metric Implementation |
| 3 | Rashi Kapoor | 0, [2], [3] | GIPSY compilation, OSS case study, Summary Analysis, Review, Quality of Classes analysis, Metric Value Analysis |
| 4 | Kanwaldeep Singh | [21], [22], [23] | Code Pro Analytix, OSS case study, Summary Analysis, Code Analysis, Logiscope, McCabe IQ, Review, Metric Implementation |
| 5 | Simar Kaur | [12], [13], [14] | MARF compilation, OSS case study, Summary Analysis, Review, Quality of Classes analysis, Metric Value Analysis |
| 6 | Shivam Patel | [7], [8], [9] | Logiscope, OSS case study, Summary Analysis, Review, McCabe IQ |
| 7 | Parth | [10], [11], [26] | OSS case study, Summary Analysis, Listing Tool Experiences, Review |
| 8 | Sagar Shukla | [4], [5], [6] | Logiscope, OSS case study, Summary Analysis, Review, McCabe IQ |